\documentclass[11pt,a4paper]{article}
\pdfoutput=1
\usepackage{caption}
\usepackage{jheppub}
\usepackage[utf8]{inputenc}
\usepackage{soul,xcolor}
\usepackage{cancel}
\usepackage{makecell}
\usepackage{diagbox}
\usepackage{makecell}
\usepackage{scalerel}
\usepackage[normalem]{ulem}
\newsavebox{\foobox}

\usepackage{multirow}
\usepackage{dcolumn}
\newcommand\Tstrut{\rule{0pt}{2.6ex}}         % = `top' strut

\newcommand\redsout{\bgroup\markoverwith{\textcolor{red}{\rule[0.5ex]{2pt}{0.8pt}}}\ULon}   % Asesh
\usepackage{ulem}
\usepackage{amsfonts}  
\usepackage{graphicx}
\usepackage{comment}
\usepackage{bm,array}
\usepackage{graphics}
\usepackage{epsf}
\usepackage{color}
\usepackage{amsmath}
\usepackage{amssymb}
\usepackage{latexsym}
\usepackage{slashed}
\usepackage{float}
\usepackage{cases}
\usepackage{multirow}
\usepackage{framed}
\usepackage{tikz}
\def\be{\begin{equation}}
\def\ee{\end{equation}}
\def\ba{\begin{array}}
\def\ea{\end{array}}

\def\alambda{A_\lambda}
\def\akappa{A_\kappa}
\def\mueff{\mu_\mathrm{eff}}
\def\beff{B_{\mathrm{eff}}}
\def\tanb{\tan\beta}

\def\sQ3{\widetilde{Q}_3}
\def\sU3{\widetilde{U}_3}
\def\sD3{\widetilde{D}_3}

\def\hsm{h_{\rm SM}}

\def\hs{h_{_S}} 
\def\as{a_{_S}}
\def\ntrli{\chi_{_i}^0}
\def\ntrlj{\chi_{_j}^0}
\def\ntrlk{\chi_{_k}^0}
\def\ntrlone{\chi_{_1}^0}
\def\ntrltwo{\chi_{_2}^0}
\def\ntrlthree{\chi_{_3}^0}
\def\ntrlfour{\chi_{_4}^0}

\def\ntrltwothree{\chi_{_{2,3}}^0}
\def\ntrlthreefour{\chi_{_{3,4}}^0}
\def\ntrltwothreefour{\chi_{_{2,3,4}}^0}

\def\charonepm{\chi_{_1}^\pm}
\def\charonemp{\chi_{_1}^\mp}

\def\mone{M_1}
\def\mtwo{M_2}

\def\mhs{m_{_{h_S}}}
\def\mhssq{m^2_{_{h_S}}}
\def\mas{m_{a_{_S}}}
\def\massq{m^2_{a_{_S}}}

\def\msinglino{m_{_{\widetilde{S}}}}
\def\msinglinosq{m^2_{_{\widetilde{S}}}}

\def\mntrlj{m_{{_{\chi}}_j^0}}

\def\mntrlone{m_{{_{\chi}}_{_1}^0}}
\def\mntrltwo{m_{{_{\chi}}_{_2}^0}}
\def\mntrlthree{m_{{_{\chi}}_{_3}^0}}
\def\mntrlfour{m_{{_{\chi}}_{_4}^0}}

\def\mcharone{m_{{_{\chi}}_{_1}^\pm}}
\def\mchartwo{m_{{_{\chi}}_{_2}^\pm}}
\def\mhsm{m_{h_{\mathrm{SM}}}}

\def\vev{{\it vev}}

\def\vu{v_u}
\def\vd{v_d}

\def\vs{v_{\!_S}}
\def\etmiss{\slashed{E}_T}

\def\nmssmtools{{\tt NMSSMTools}}
\def\checkmate{{\tt CheckMATE}}
\def\smodels{{\tt SModelS}}
\def\micromegas{{\tt micrOMEGAs}}
\def\higgsbounds{{\tt HiggsBounds}}
\def\higgssignals{{\tt HiggsSignals}}

\def\pythia8{{\tt PYTHIA8}}

\def\z3nmssm{$Z_3$-NMSSM}
\newcommand{\beq}{\begin{equation}}
\newcommand{\eeq}{\end{equation}}
\newcommand{\bea}{\begin{eqnarray}}
\newcommand{\eea}{\end{eqnarray}}
\title{Dark Matter searches with photons at the LHC}
\author[a]{Subhojit Roy}
\author[a,b,c]{and Carlos E.M. Wagner}
\affiliation[a]{HEP Division, Argonne National Laboratory, 9700 Cass Ave., Argonne, IL 60439, USA}
\affiliation[b]{Enrico Fermi Institute, Physics Department, University of Chicago, Chicago, IL 60637, USA}
\affiliation[c]{Kavli Institute for Cosmological Physics, University of Chicago, Chicago, IL 60637, USA}
\emailAdd{sroy@anl.gov, cwagner@anl.gov}%,
\preprint{EFI 24-1} 
\abstract{ 
We unveil blind spot regions in dark matter (DM) direct detection (DMDD), for weakly interacting massive particles with a mass around a few hundred~GeV that may reveal interesting photon signals at the LHC. 
We explore a scenario where the DM primarily originates from the singlet sector within the $Z_3$-symmetric Next-to-Minimal Supersymmetric Standard Model (NMSSM). A novel DMDD spin-independent blind spot condition is revealed for singlino-dominated DM, in cases where the mass parameters of the higgsino and the singlino-dominated lightest supersymmetric particle (LSP) exhibit opposite relative signs (i.e., $\kappa < 0$), emphasizing the role of nearby bino and higgsino-like states in tempering the singlino-dominated LSP. Additionally, proximate bino and/or higgsino states can act as co-annihilation partner(s) for singlino-dominated DM, ensuring agreement with the observed relic abundance of DM. Remarkably, in scenarios involving singlino-higgsino co-annihilation, higgsino-like neutralinos can distinctly favor radiative decay modes into the singlino-dominated LSP and a photon, as opposed to decays into leptons/hadrons. In exploring this region of parameter space within the singlino-higgsino compressed scenario, we study the signal associated with at least one relatively soft photon alongside a lepton, accompanied by substantial missing transverse energy ($\etmiss$) and a hard initial state radiation jet at the LHC. In the context of singlino-bino co-annihilation, the bino state, as the next-to-LSP, exhibits significant radiative decay into a soft photon and the LSP, enabling the possible exploration at the LHC through the triggering of this soft photon alongside large $\etmiss$ and relatively hard leptons/jets resulting from the decay of heavier higgsino-like states.
}
\begin{document}
\maketitle
%
%%%%%%%%%%%%%%%%%%%%%%%%%%%%%%%%%%
\section{Introduction}
\label{sec:Intro}
%%%%%%%%%%%%%%%%%%%%
%
Dark matter (DM) is a fundamental component of the present universe, playing a vital role in the formation of structures and providing explanations for phenomena like the rotation patterns of galaxies and various observations in astrophysics and cosmology. 
In recent times, the discovery of the Higgs Boson ($\hsm$) at the Large Hadron Collider (LHC) around 125~GeV~\cite{ATLAS:2012yve, CMS:2012qbp} was a breakthrough moment in our understanding of the laws of nature, completing the particle content of the Standard Model (SM). 
However, following this discovery, in the present understanding of particle physics, the most crucial and compelling questions are the nature of the DM and the origin of the weak scale, which is not protected under radiative corrections caused by the heavy particles that interact with the Higgs boson.
Low-scale supersymmetry (SUSY)~\cite{Nilles:1983ge,Haber:1984rc,Martin:1997ns} emerges as a highly versatile and among the most popular beyond-the-Standard-Model (BSM) scenarios since it not only offers a viable candidate for DM~\cite{Goldberg:1983nd, Ellis:1983ew} and solution to the stability problem of the electroweak scale but also addresses some other theoretical issues of the SM such as the instability of the electroweak vacuum and the (non)unification of gauge couplings.

In such a scenario, the lightest supersymmetric Particle (LSP) can potentially serve as a suitable candidate for DM  when a well-known discrete symmetry called $R$-parity is conserved.
In this study, we investigate the scenario within the $Z_3$-symmetric Next-to-Minimal Supersymmetric Standard Model (NMSSM)~\cite{Ellwanger:2009dp} where the LSP is singlino-dominated neutralino, i.e., originating primarily from the singlet sector. As it is neutral, stable, and exhibits weak interactions, it behaves as a Weakly Interacting Massive Particle (WIMP)-type DM candidate.
Such a singlino-dominated LSP is tempered by the other nearby neutralinos~\cite{Arkani-Hamed:2006wnf}, which are the superpartners of the electroweak gauge and Higgs bosons, i.e., the bino, wino and higgsinos.
To avoid over-closing the universe and comply with the Planck experiment~\cite{Planck:2015fie, Planck:2018vyg} reported DM relic density, sufficient annihilation processes of the singlino-dominated DM in the early universe is required.
A pair of singlino-dominated DMs can annihilate via the usual resonant $s$-channel process exchanging the Z-boson, $\hsm$, the other doublet-like ($H, A$) and singlet-like ($\hs, \as$) scalars and through
$t$-channel processes via the exchange of a chargino or neutralino. It can also co-annihilate with other nearby neutralino and chargino states. 
Furthermore, since the singlino-dominated DM can have some bino mixing, DM co-annihilation with the superpartners of leptons (sleptons) is also possible. But for the present
work, such a process is not possible as we have fixed all the slepton masses at a few TeV.

In recent years, numerous dedicated searches have been conducted to detect DM through various experiments focused on DM direct detection (DMDD) and in colliders such as the LHC.
%The continuously improving sensitivity of dark matter direct detection experiments
The stringent and improved upper bounds on spin-independent (SI)
and spin-dependent (SD) DM-nucleus (elastic) scattering cross-sections that are
routinely arriving from various DMDD experiments~\cite{XENON:2019rxp, XENON:2023cxc, PandaX-4T:2021bab, LZ:2022lsv, PICO:2017tgi, IceCube:2021xzo, Fermi-LAT:2017bpc, MAGIC:2016xys}, 
put constraints on the interaction rates of a pair of DMs with $\hsm$ and other Higgs bosons of the model and the gauge boson $Z$.  For values of the supersymmetric particle masses of the order of the weak scale,
these limits necessitate the theoretical framework to be in close proximity to blind spots.
In this work, an involved set of DMDD-SI and -SD blind spot conditions for the singlino-dominated DM are derived by considering $(4\times4)$ bino-higgsino-singlino neutralino sector. We consider the wino to be decoupled.

It is well known that at the minimal mixing between the SM-like Higgs boson and the other Higgs bosons and considering only the singlino-higgsino neutralino sector, DMDD-SI coupling blind spot condition arises exclusively when the ratio $\mntrlone/\mueff > 0$, i.e., $\kappa > 0$~\cite{Badziak:2015exr, Baum:2017enm, Cheung:2014lqa, Badziak:2017uto, Ellwanger:2004xm}.
In this blind spot scenario, the effect of the $\lambda$ governed singlino-higgsino-Higgs interaction term on the LSP becomes significantly suppressed.
In this study, we demonstrate that a new blind spot condition can emerge for singlino-dominated DM due to the tempering with the bino and higgsino-like neutralinos. This occurs when singlino mass parameter $2 \kappa \mueff/\lambda$ and $\mueff$ exhibit an opposite relative sign, corresponding, for $\lambda > 0$, to the $\kappa < 0$ region in the parameter space. This unique scenario results from a cancellation between the effects of the $g_1$ governed gaugino-higgsino-Higgs interaction term and the $\lambda$ proportional singlino-higgsino-Higgs interaction term on the LSP.
Hence, in general, the blind spot condition arises when there is the same (or opposite) relative sign between $\mueff$ and $M_1$ for $\kappa < 0$ ($> 0$).
This opens up a new region of parameter space characterized by a smaller DMDD-SI cross-section for the singlino-dominated DM.

This newly found DMDD-SI blind spot region of parameter space of $\kappa < 0$ can exhibit an interesting link on the observed discrepancy between the experimental observations (from Fermilab and BNL) and the SM prediction of the anomalous muon magnetic moment~\cite{Muong-2:2006rrc, Muong-2:2021ojo}.
Since the experimentally measured value is larger than the SM prediction, a positive contribution from new physics is required to explain this discrepancy. Note that for large values of the Wino mass $M_2$, assumed in this work, the DMDD-SI blind spot condition for $\kappa < 0$ demands the same relative sign between $M_1$ and $\mueff$, which also provides a positive contribution from the Bino-smuon loop to the anomalous muon magnetic moment that can explain the observed discrepancy in the presence of light smuons.
It is important to remember that this discrepancy is based on the theoretical estimation utilizing cross-section measurements data from ($e^+ e^- \to {\rm hadrons}$) to determine the hadronic vacuum polarization (HVP) contribution through dispersion relations~\cite{Aoyama:2020ynm}.
Presently, a disagreement exists in the assessment of the HVP contribution between lattice calculations~\cite{Borsanyi:2020mff} and this data-driven determination (for a discussion of this issue, see reference~\cite{Coyle:2023nmi}). 
As this disparity remains unresolved and awaits further scrutiny, our focus in this work is not directed towards investigating this phenomenon in the NMSSM framework.
 Consequently, we treat smuons as heavy in our analysis. However, if this discrepancy persists, it would be intriguing to conduct a comprehensive exploration, incorporating light smuons, within the context of DM phenomenology and the anomalous muon magnetic moment in NMSSM.

In recent years, various searches have been conducted to discover new particles at the LHC.
Searches for electroweakinos are conducted from the direct pair-production of the wino-like charginos and neutralinos ($\charonepm \ntrltwo$) at the LHC~\cite{Liu:2020ctf} with the assumption that 
they eventually decay into two modes: $W^{\pm}Z\ntrlone$ and $W^{\pm}\hsm\ntrlone$ and the LSP ($\ntrlone$) is bino-like. Such decay modes are studied considering final states  characterized by $3\ell + \etmiss$~\cite{ATLAS:2020ckz, CMS:2017moi, CMS:2018szt, ATLAS:2018ojr, ATLAS:2018eui, 
ATLAS:2019wgx, CMS:2021edw}, $1\ell + 2b$-${\rm jet} + \etmiss$~\cite{CMS:2017moi, CMS:2018szt, ATLAS:2018qmw, CMS:2019pov, 
ATLAS:2020pgy, ATLAS:2020ckz} and $2\ell + 2$-${\rm jet} + \etmiss$~\cite{CMS:2017moi, CMS:2018szt, ATLAS:2018ojr, ATLAS:2018eui, CMS:2021edw}. The pair production of the charginos, i.e., $\charonepm \charonemp$, is investigated by considering their decay into the $W^{\pm}W^{\mp}\ntrlone$ mode in the final state of $2\ell + \etmiss$~\cite{ATLAS:2019lff, CMS:2018xqw}.
Dedicated searches are conducted considering final states comprising soft multi-leptons/jets, $\etmiss$, and possible ISR jets to explore the compressed region of parameter space.~~\cite{ATLAS:2019lng, ATLAS:2018ojr,ATLAS:2019lff,CMS:2020bfa,ATLAS:2021moa,CMS:2021cox,CMS:2021few,ATLAS:2021yqv,ATLAS:2022zwa,CMS:2022sfi,CMS:2019san}.
 These searches are motivated assuming $\ntrltwo \rightarrow \ntrlone f \bar{f}$ decay mode where the pair of soft leptons/jets ($f \bar{f}$) come from the off-shell $Z$ and $\hsm$.
 However, null results from these
searches of electroweakinos at the LHC put constraints on their masses.
 In this work, we focus on the compressed region of parameter space where the next-to LSP (NLSP) is either higgsino-like or bino-like neutralino that exhibits substantial radiative decay modes characterized by larger branching fraction
to a photon and the singlino-dominated LSP ($\ntrltwo \rightarrow \ntrlone \gamma$).
Furthermore, a novel region of parameter space is discussed where both higgsino-like neutralino states ($\ntrltwo$ and $\ntrlthree$) 
features enhanced branching fraction in the $\ntrltwo \rightarrow \ntrlone \gamma$ and $\ntrlthree \rightarrow \ntrltwo \gamma$ decay modes.
These distinct decay patterns may lead to intriguing collider signatures at the LHC.
Note that certain decay channels among these have not been explored in previous LHC searches, resulting in a relaxation of mass bounds for electroweakinos.
This region of parameter space is primarily motivated by the DMDD blind spots, where the DMDD-SI scattering cross-section either stays below the latest bounds from the latest DMDD experiments or even falls below the irreducible neutrino background.
Therefore, it is important to explore these regions of the DM parameter space at the LHC.

In the scenario of singlino-higgsino co-annihilation, the small mass gap between higgsino-like states ($\ntrltwothree$) and the singlino-dominated LSP leads to soft decaying objects from higgsino-like neutralinos and charginos, arising from their associated production processes $pp \rightarrow \ntrltwothree \charonepm$. Therefore, suppressing the SM backgrounds becomes difficult. 
In this work, we discuss, illustrating various differential distributions of kinematic variables, that this scenario can be addressed by analyzing the $pp \rightarrow \ntrltwothree \charonepm$ process with an initial-state radiation jet~\cite{Baum:2023inl}, considering a signal with hard mono-jet, significant missing energy, and at least one photon.
Furthermore, we show that the presence of a bino-like fourth neutralino state indirectly influences both collider and DMDD phenomena.  
It tempers the neutralinos in such a way that the DMDD-SI cross-section of the singlino-dominated DM remains small, and higgsino-like neutralinos feature a larger branching fraction in the radiative photonic decay mode.

In the singlino-bino co-annihilation scenario, the bino-like NLSP can undergo a significant radiative decay~\cite{Domingo:2018ykx, Baer:2005jq}, emitting a photon along with the singlino-dominated DM.
In this scenario, a bino-like NLSP can emerge with a significant boost resulting from the decay of the heavier higgsino-like neutralinos and charginos. Such a boost can prevent the photon, originating from the decay of the NLSP, from remaining soft.
The leptons/ jets originating from the on-shell $Z$, $\hsm$ and $W^{\pm}$-bosons, that are coming from the cascades of the heavier higgsino-like states, are relatively hard. 
The cascades of the produced higgsino-like $\ntrlthreefour$ and $\charonepm$ via $\ntrltwo$ could lead to some final states such as $3\ell + \geq 1\gamma + \etmiss$ or $1\ell + 2b + \geq 1\gamma + \etmiss$ at the LHC.
Note that in some situations, the cascades of $\charonepm$ and $\ntrlthreefour$ via bino-like $\ntrltwo$ can lead to similar final states as when these higgsino-like states directly decay to the singlino-dominated LSP, if the photon coming from $\ntrltwo$ remains soft and undetected.
Therefore, observing excess over the SM backgrounds in the signals with and without soft photons with  $3\ell + \etmiss$ and $1\ell + 2b  + \etmiss$ at the HL-LHC could promptly point back to the scenario that includes singlino-bino co-annihilation with relatively heavier higgsinos.
Furthermore, utilizing \checkmate~\cite{Dercks:2016npn}, we point out that the singlino-bino co-annihilation region of parameter space can be constrained from an analysis~\cite{ATLAS:2020qlk} conducted by ATLAS.
The analysis primarily aims to search for neutralinos/charginos, considering a pair of photons originating from the decay of the on-shell $\hsm$, which arises from the decay of a heavier neutralino.
As the analysis involves photons, some of the signal regions can overlap with this specific singlino-bino co-annihilation region of the parameter space of NMSSM.

The paper is organized as follows. In section~\ref{sec:scenario}, we describe the theoretical framework by presenting the superpotential of NMSSM along with its key aspects, including the electroweakino and scalar (Higgs) sectors. The interplay of the Higgs sector and the electroweakino sector,  involving correlations in their interactions and masses, is subsequently explored, considering LSP to be singlino-dominated. Furthermore, the phenomenology of the singlino-dominated DM relic abundance and various DMDD-SI and DMDD-SD blind spot conditions considering the singlino-higgsino-bino $(4 \times 4)$ neutralino system are discussed.
In section~\ref{constraints}, we discuss various constraints which are relevant to the present work, coming from both the dark sector and the LHC.
The motivated region of the NMSSM parameter space of the present work is discussed in section~\ref{subsec:motivatedregion}.
Furthermore, in section~\ref{DMsearchwithphoton}, we discuss the searches of singlino-dominated DM with photons at the LHC. We present a few benchmark scenarios that feature photons from the cascades of the heavier electroweakinos to the DM and capture the salient aspects of the present work. Further, a brief observation is made about the new search channels of those benchmark points at the LHC. Finally, we conclude in section~\ref{conclusion}.
%
%%%%%%%%%%%%%%%%%%%%%%%%%%%%%%%%%%
\section{The theoretical scenario}
\label{sec:scenario}
%%%%%%%%%%%%%%%%%%%%
The superpotential of the $Z_3$-symmetric NMSSM is given by~\cite{Ellwanger:2009dp}
\beq
{\cal W}= {\cal W}_\mathrm{MSSM}|_{\mu=0} + \lambda \widehat{S}
\widehat{H}_u \cdot \widehat{H}_d
        + {\kappa \over 3} \widehat{S}^3 \, ,
\label{eqn:superpot}
\eeq
where ${\cal W}_\mathrm{MSSM}|_{\mu=0}$ is the MSSM superpotential including the Yukawa interactions of the Higgs doublet superfields with the SM quark and leptons superfields, but with no
higgsino mass term (known as $\mu$-term), $\widehat{H}_u, \widehat{H}_d$ and $\widehat{S}$ are the $SU(2)$ Higgs doublet superfields of the MSSM and the gauge
singlet superfield of the $Z_3$-symmetric NMSSM, respectively, and `$\lambda$' and `$\kappa$' are dimensionless coupling constants. Assuming the conservation of $R$-parity,  this is the most general $Z_3$-symmetry superpotential.  Adding the extra superfield $\widehat{S}$ to the MSSM solves the so-called $\mu$-problem~\cite{Kim:1983dt} when it gets non-zero vacuum expectation value ~(\vev)~$\vs$ and generates dynamically an effective $\mu$-term given by $\mueff=\lambda \vs$ from the second term in eq.~(\ref{eqn:superpot}). The soft SUSY-breaking Lagrangian is given by
\beq
-\mathcal{L}^{\rm soft}= -\mathcal{L_{\rm MSSM}^{\rm soft}}|_{B\mu=0}+ m_{S}^2
|S|^2 + (
\lambda A_{\lambda} S H_u\cdot H_d
+ \frac{\kappa}{3}  A_{\kappa} S^3 + {\rm h.c.}) \,,
\label{eqn:lagrangian}
\eeq
where $m_S$ is the soft SUSY-breaking mass
of the singlet scalar field, `$S$' and $\alambda$ and $\akappa$ are the NSSM-specific trilinear soft couplings with mass dimension one, and $\mathcal{L_{\rm MSSM}^{\rm soft}}|_{B\mu=0}$ includes no $H_u H_d$ Higgs bilinear soft supersymmetry breaking term. In the next subsections, we briefly discuss the electroweakino sector and the Higgs sector and their interactions.
%%%%%%%%%%%%%%%%%%%%%%%%%%%%%
\subsection{The electroweakino sector}
\label{subsec:ewinos}
%%%%%%%%%%%%%%%%%%%%%
The electroweakino sector contains the neutralinos and the charginos. NMSSM has one extra neutralino known as singlino ($\widetilde{S}$) state coming from the gauge-singlet superfield appearing in the 
superpotential of NMSSM in eq.~(\ref{eqn:superpot}) compared to the MSSM. Thus, the symmetric $(5\times 5)$ neutralino mass matrix, in the basis $\psi^0=\{\widetilde{B},~\widetilde{W}^0, ~\widetilde{H}_d^0,
~\widetilde{H}_u^0, ~\widetilde{S}\}$, is given by~\cite{Ellwanger:2009dp}
\beq
\label{eqn:mneut}
{\cal M}_0 =
\left( \begin{array}{ccccc}
\mone & 0 & -\dfrac{g_1 \vd}{\sqrt{2}} & \dfrac{g_1 \vu}{\sqrt{2}} & 0 \\[0.4cm]
\ldots & \mtwo & \dfrac{g_2 \vd}{\sqrt{2}} & -\dfrac{g_2 \vu}{\sqrt{2}} & 0 \\
\ldots & \ldots & 0 & -\mueff & -\lambda \vu \\
\ldots & \ldots & \ldots & 0 & -\lambda \vd \\
\ldots & \ldots & \ldots & \ldots & 2 \kappa \vs
\end{array} \right) \,,
\eeq
where 
%$g^2=(g_1^2+g_2^2)/2$, 
$g_1$ and $g_2$ are the $U(1)_Y$ and 
$SU(2)_L$ gauge couplings, respectively, and $\vu=v\sin\beta$, $\vd=v\cos\beta$ such that
$v^2=\vu^2+\vd^2 \approx (174$~GeV)$^2$ and $\tan\beta=\vu/ \vd$.
$\mone$ ($\mtwo$) is the soft SUSY-breaking masses for the $U(1)_Y$ ($SU(2)_L$) gaugino, known as bino (wino).
The (5,5) element of the neutralino mass-matrix in eq.~(\ref{eqn:mneut}) is the singlino mass term, $\msinglino = 2\kappa \vs$. 

The symmetric matrix ${\cal M}_0$ in the absence of $CP$-violation can be diagonalized by an orthogonal $(5 \times 5)$ matrix `$N$',
i.e.,
\bea
\label{eqn:diagonalise-1}
N {\cal M}_0 N^T = {\cal M}_D = {\rm diag}(m_{{_{\chi}}_{_1}^0},m_{{_{\chi}}_{_2}^0},m_{{_{\chi}}_{_3}^0},m_{{_{\chi}}_{_4}^0},m_{{_{\chi}}_{_5}^0})  \, . 
\eea
The neutralino mass-eigenstates ($\ntrli$) are represented in terms of the weak eigenstates by ($\psi_j^0$)
\beq
\ntrli = N_{ij} \psi_j^0 \,,
\label{eqn:diagN2}
\eeq
where $i$ and $j$ run from 1 to 5. This study considers a scenario in which the wino mass value is decoupled from the other neutralino states. As a result, the heaviest neutralino state becomes a wino-like state and $N_{52} \sim 1$. The $(5\times 5)$ symmetric neutralino mass-matrix of eq.~(\ref{eqn:mneut}) effectively reduces to a $(4\times 4)$ matrix contained the bino, the higgsinos and the singlino states at the decoupled wino limit. The relations among these components of a $j$-th neutralino state are given by:
\vspace{-0.25cm}
\bea
\label{eqn:N15-N11relation-1}
{N_{j5} \over N_{j1}}
  &=& {-{g^2_1 \over 2} v^2(\mueff \sin 2\beta + \mntrlj) + (M_1 - \mntrlj)(\mueff^2 - \mntrlj^2) \over {g_1 \over \sqrt{2}}\lambda\, \mueff\,(v^2_u -v^2_d)} \, , \\
\label{eqn:N13-N11relation-1}
{N_{j3} \over N_{j1}}
  &=& {{g^2_1 \over 2}v^2 v_u + (M_1 - \mntrlj)(v_u \,\mntrlj - v_d \,\mueff) \over {g_1 \over \sqrt{2}} \,\mueff \,(v^2_u -v^2_d)} \, ,\\
\label{eqn:N14-N11relation-1}
{N_{j4} \over N_{j1}}
  &=& {{g^2_1 \over 2}v^2 v_d  + (M_1 - \mntrlj)(v_d \,\mntrlj - v_u\, \mueff) \over {g_1 \over \sqrt{2}} \,\mueff \,(v^2_u -v^2_d)} \, ,\quad  
\eea
where $N_{j1}$, $N_{j5}$ are the bino and the singlino admixtures, respectively, and  $N_{j3}$, $N_{j4}$ are the higgsino components in the $j$-th neutralino state. 
Various components can be obtained from the unitarity relation, i.e., $N^2_{j1} + N^2_{j3} + N^2_{j4} + N^2_{j5} \simeq 1$
For example, the bino component in the LSP is given by,
\bea
\label{eqn:n^2j5-value}
N^2_{11} = {Z^2 \over I(M_1 - \mntrlone)^2 \, (\mueff^2 - \mntrlone^2)} \,,
\eea
where
\vspace{-0.5cm}
\bea
\label{eqn:Z^2-value}
Z =  {g_1 \over \sqrt{2}}\, \lambda\,   \mueff\, (v^2_u -v^2_d),
\eea
and
\bea
\label{eqn:I-value}
I &=& \mueff^2 - \mntrlone^2 +  (\lambda v)^2{\mntrlone^2 + \mueff^2 - 2 \mntrlone \mueff \sin 2\beta \over {\mueff^2 - \mntrlone^2}} + {\lambda^2 g^2_1 v^4 (\mntrlone - \mueff \sin 2\beta) \over (M_1 - \mntrlone) \, (\mueff^2 - \mntrlone^2)} \nonumber \\ 
 &-& {g_1^2 v^2 (\mueff \sin 2\beta + \mntrlone) \over {M_1 - \mntrlone}} + {Z^2 + g_1^4 v^4\big[\lambda^2_1 v^2 + (\mueff \sin 2\beta + \mntrlone)^2 \big] \over (M_1 - \mntrlone)^2 \, (\mueff^2 - \mntrlone^2)} \,. \hskip 35pt \quad
\eea
%It is straightforward to find 
 This is an extension of the relations obtained
for the (3$\times$3) singlino-higgsino and bino-higgsino systems~\cite{Baum:2017enm, Abdallah:2020yag} from eqs. (\ref{eqn:N15-N11relation-1}), (\ref{eqn:N13-N11relation-1}), (\ref{eqn:N14-N11relation-1}) via decoupling $M_1$ and $\msinglino$, respectively.
In this work, we will utilize these relations provided in eqs. (\ref{eqn:N15-N11relation-1}) to (\ref{eqn:I-value}) to calculate various couplings of a pair of singlino-dominated LSPs with other particles.
%Now write the neutralino  various relations

The chargino sector in the NMSSM is exactly similar to the MSSM but for $\mu$ $\rightarrow$ $\mueff$. The ($2 \times 2$) chargino mass matrix
in the bases
$\psi^+ = \{ -i \widetilde{W}^+, \, \widetilde{H}_u^+ \}$ and
$\psi^- = \{ -i \widetilde{W}^-, \, \widetilde{H}_d^- \}$, is given by~\cite{Ellwanger:2009dp}
\beq
{\cal M}_C = \left( \begin{array}{cc}
                    \mtwo   & \quad  g_2 \vu \\
                 g_2 \vd  & \quad \mueff 
             \end{array} \right) .
\eeq
Similar to MSSM, ${\cal M}_C$ can be diagonalized by two ($2 \times 2$) 
unitary matrices `$U$' and `$V$':
\beq
U^* {\cal M}_C V^\dagger = \mathrm{diag} (\mcharone , \mchartwo) \; ; \quad
\mathrm{with} \;\;  \mcharone < \mchartwo  \,.
\label{eqn:uvmatrix}
\eeq
%
%%%%%%%%%%%%%%%%%%%%%%%%%%%%%%%%%%%%%%
\subsection{The Higgs sector and the interactions with electroweakinos}
\label{subsec:higgs}
%%%%%%%%%%%%%%%%%%%%
%
The soft
Lagrangian for the NMSSM Higgs sector  from eq.~(\ref{eqn:lagrangian}) is given below:
\beq\label{2.5e}
-{\cal L}^\mathrm{soft} \supset
m_{H_u}^2 |H_u|^2 + m_{H_d}^2 | H_d |^2 
+ m_{S}^2 |S|^2
+\left(\lambda A_\lambda S H_u \cdot H_d + \frac{\kappa}{3}  A_\kappa
S^3 
+ \mathrm{h.c.}\right) .
\eeq
The neutral Higgs fields $H_d^0$, $H_u^0$ and $S$ can be expanded around their real  \vev's  i.e.,  $v_d$, $v_u$ and $\vs$. Thus,
\beq\label{2.10e}
H_d^0 = v_d + \frac{H_{dR} + iH_{dI}}{\sqrt{2}} , \quad
H_u^0 = v_u + \frac{H_{uR} + iH_{uI}}{\sqrt{2}} , \quad
S = \vs + \frac{S_R + iS_I}{\sqrt{2}},
\eeq
where `$R$' and `$I$' correspond to the $CP$-even and the $CP$-odd states, respectively. The $CP$-even scalars symmetric squared mass matrix (${\cal M}_S^2$) in the basis
$H_{jR}=\{H_{dR}, H_{uR}, S_R\}$ is given by~\cite{Ellwanger:2009dp}
\beq
{\small{
{\cal M}_S^2 =
\left( \begin{array}{ccc}
  g^2 \vd^2 + \mueff \beff \,\tanb
& \:\; (2\lambda^2 - g^2) \vu \vd - \mueff \beff
& \:\; \lambda (2 \mueff \, \vd - (\beff + \kappa \vs) \vu) \\[0.2cm]
 \ldots
& \:\; g^2 \vu^2 + \mueff \beff /\tanb
& \:\; \lambda (2 \mueff \, \vu - (\beff + \kappa \vs)v_d) \\[0.2cm]
 \ldots
 &  \ldots
& \:\:\, \lambda \alambda  \frac{\vu \vd}{\vs} + \kappa \vs (\akappa + 4\kappa \vs)
\label{eqn:cp-even-matrix} 
\end{array} \right),}
}
\eeq
where $B_{\mathrm{eff}}=\alambda+\kappa\vs$ and $g^2 = (g_1^2 + g_2^2)/2$.
The $CP$-even Higgs bosons ($h_i$) mass eigenstates are then given by
\bea
h_i &=& S_{ij} H_{jR}, \qquad  
\mathrm{with} \quad {i,j=1,2,3}\, ,
\label{eqn:cp-even-scalar-physical-states}
\eea
where the matrix `$S$' diagonalizes ${\cal M}_{S}^2$.
The interaction coupling of the $CP$-even scalars ($h_i$), a pair of neutralinos $\ntrlj$ and $\ntrlk$ are given by~\cite{Ellwanger:2009dp, Ellwanger:2004xm}
\bea
g_{_{h_i \ntrlj \ntrlk}} &=& {\lambda \over \sqrt{2}}
(S_{i1} \Pi^{45}_{jk} + S_{i2} \Pi^{35}_{jk} + S_{i3} \Pi^{34}_{jk})
 - \sqrt{2} \, \kappa S_{i3} N_{j5} N_{k5} \nonumber \\
&+& {g_1 \over 2} (S_{i1} \Pi^{13}_{jk} - S_{i2} \Pi^{14}_{jk})
 -  {g_2 \over 2} (S_{i1} \Pi^{23}_{jk} - S_{i2} \Pi^{24}_{jk}),
\label{eqn:hinjnk1}
\eea
where $\Pi^{ab}_{jk} = N_{ja} N_{kb} + N_{jb} N_{ka}$.

In a more convenient rotated basis, ($\hat{h}, \widehat{H}, \hat{s}$)~\cite{Miller:2003ay, Badziak:2015exr} where $\hat{h} = H_{dR} \cos{\beta} + H_{uR} \sin{\beta}$, $\widehat{H} = H_{dR} \sin{\beta} - H_{uR} \cos{\beta}$ and $\hat{s} = S_{R}$, $\hat{h}$ mimics the SM Higgs field where as $\widehat{H}$ resembles the MSSM heavy doublet-like $CP$-even Higgs. The physical $CP$-even scalar states are given by 
\bea
h_i &=& E_{h_i \hat{h}} \hat{h} + E_{h_i \widehat{H}} \widehat{H} + E_{h_i \hat{s}} \hat{s} \,,
\label{eqn:hiEhhat}
\eea
where $E_{ab}$ is the diagonalizing matrix of the mass-squared
matrix for the $CP$-even scalars in the rotated basis.
In this rotated basis ($\hat{h}, \widehat{H}, \hat{s}$) eq.~(\ref{eqn:hinjnk1}) reduces to
\bea
\label{eqn:hinjnk-reduced-without-approximation}
g_{_{h_i \ntrlj \ntrlk}}
 & = &
\Bigg[{\lambda \over \sqrt{2} } \big[ E_{h_i \hat{h}} N_{j5} (N_{k3} \sin\beta + N_{k4} \cos\beta)
 + E_{h_i \hat{H}} N_{j5} (N_{k4} \sin\beta - N_{k3} \cos\beta)  \nonumber \\
 & & \hskip 25pt + \; E_{h_i \hat{s}} (N_{j3}N_{k4} - {\kappa \over \lambda} N_{j5}N_{k5})\big] + {1 \over 2}\big[g_{1}N_{j1} - g_{2}N_{j2}\big] \big[E_{h_i \hat{h}} (N_{k3} \cos\beta - N_{k4} \sin\beta) \nonumber \\
 & & \hskip 25pt + E_{h_i \hat{H}} (N_{k3} \sin\beta + N_{k4} \cos\beta)\big]\Bigg] + \Bigg[j \longleftrightarrow k \Bigg]\, .
\eea
For no mixing between the singlet and the doublet scalars ($E_{\hsm \hat{s}}, E_{H \hat{s}} \sim 0$) and between the doublet Higgs states (i.e., $E_{\hsm \hat{h}}, E_{H \hat{H}} \sim 1$ and $E_{\hsm \hat{H}} \sim 0$)~\cite{Badziak:2015exr}, the coupling of a pair of LSPs with the SM-like Higgs boson is given by~\cite{Abdallah:2020yag}, 
\bea
\label{eqn:hin1n1-2}
g_{_{\hsm \ntrlone \ntrlone}}
 &\simeq&
\sqrt{2} \lambda \, (N_{13} \sin\beta + N_{14} \cos\beta)N_{15} + g_{1} (N_{13} \cos\beta - N_{14} \sin\beta)N_{11},
\eea
where the $g_1$-proportional second term in (\ref{eqn:hin1n1-2}) has pure MSSM origin (gaugino-higgsino-Higgs interaction) and the first $\lambda$-proportional term is only possible in the NMSSM due to the higgsino-singlino-Higgs interaction. Using eqs. (\ref{eqn:N15-N11relation-1}) to (\ref{eqn:I-value}), eq.~(\ref{eqn:hin1n1-2}) reduces to
\bea
\label{eqn:hin1n1-3}
\hskip -30pt
g_{_{\hsm \ntrlone \ntrlone}}
 &\simeq&
{\sqrt{2}\lambda^2 v \over I} \Bigg[ \mntrlone - \mueff \sin 2\beta + {{g_1^2 v^2} \over {M_1 - \mntrlone}} + {g_1^4 v^4  \over 4}{\mntrlone + \mueff \sin 2\beta \over (M_1 - \mntrlone)^2 \, (\mueff^2 - \mntrlone^2)} \Bigg] \,.
\eea
This coupling holds particular importance as it directly influences the spin-independent (SI) scattering cross-section for the direct detection of DM. Subsequently, we will utilize it to obtain the ``coupling blind spot" for the singlino-dominated DMDD-SI cross-section.

On the other hand, the $CP$-odd scalars (3 $\times$ 3) symmetric squared mass matrix in the basis
$H_{jI}=\{H_{dI}, H_{uI}, S_I\}$ is given by~\cite{Ellwanger:2009dp}
\beq
{\small{
{\cal M'}_{P}^2 =
\left( \begin{array}{ccc}
  \mueff \beff \,\tanb
& \mueff \beff
& \lambda \vu (\alambda - 2\kappa \vs) \\[0.2cm]
  \ldots
& \mueff \beff /\tanb
& \lambda \vd (\alambda - 2\kappa \vs) \\[0.2cm]
  \ldots
&  \ldots
& \lambda (\beff + 3 \kappa \vs) \frac{\vu \vd}{\vs} - 3 \kappa \akappa \vs
\label{eqn:cp-odd-matrixp}
\end{array} \right),}
}
\eeq
Where $\beff = \alambda v_s$. Similar to the $CP$-even Higgs sector, working on the rotated basis $\{A, S_I \}$ where $A=\cos\beta H_{uI} + \sin\beta H_{dI}$, dropping the massless Nambu-Goldstone mode the (2 $\times$ 2) $CP$-odd (pseudoscalar) Higgs boson squared mass matrix is given by,
\beq\label{eqn:cp-odd-matrix}
{\small
{\cal M}_P^2 =
\left( \begin{array}{cc}
    m_A^2
~&~ \lambda (\alambda - 2\kappa \vs)\, v \\[0.2cm]
    \lambda (\alambda - 2\kappa \vs)\, v
~&~ \lambda (\alambda + 4\kappa \vs)\frac{v_u v_d}{\vs} -3\kappa \akappa \, \vs  
\end{array} \right),
}
\eeq
where,  $m_A^2= 2 \mueff \beff / \sin2\beta$ is the MSSM-like $CP$-odd Higgs squared mass. The $CP$-odd (pseudoscalar, $a_k$) mass eigenstates, in terms of ${\cal M'}_{P}^2$ diagonalization matrix `${\cal O}$', are given~by
\bea
a_k = {\cal O}_{kA} A + {\cal O}_{kS_I} S_I, 
\qquad \mathrm{with} \quad {k=1,2},
\label{eqn:cp-odd-scalar-physical-states}
\eea
The relations among the elements of the `${\cal O}$' and `$P$' which diagonalizes ${\cal M'}_{P}^2$ are given by~\cite{Ellwanger:2009dp}
\beq
 P_{i1} = \sin\beta \, {\cal O}_{iA} \,, \qquad 
 P_{i2} = \cos\beta\,  {\cal O}_{iA}\,, \qquad
 P_{i3} = {\cal O}_{iS_I} \,.
\eeq
The interaction coupling of the $CP$-odd scalars, $a_i$ and a pair of neutralinos, $\ntrlj$ and $\ntrlk$ is given by,
\bea
\label{eqn:ainjnk-reduced-without-approximation}
g_{_{a_i \ntrlj \ntrlk}}
 & = &
\Bigg[i\Big({\lambda \over \sqrt{2}} \big[{\cal O}_{iA} N_{j5} (N_{k4} \sin\beta + N_{k3} \cos\beta) + {\cal O}_{iS_I} (N_{j3}N_{k4} - {\kappa \over \lambda} N_{j5}N_{k5})\big]  \nonumber \\
 & & \hskip 25pt - \; {1 \over 2}\big[g_{1}N_{j1} - g_{2}N_{j2}\big] \big[{\cal O}_{iA} (N_{k3} \sin\beta - N_{k4} \cos\beta)\big]\Big)\Bigg] + \Bigg[j \longleftrightarrow k \Bigg] . \quad
\eea
For singlet-dominated $CP$-odd Higgs boson ${\cal O}_{iS_I} \sim 1$ and ${\cal O}_{iA} \sim 0$. This indicates that the $CP$-odd singlet Higgs boson interaction with a pair of singlino-dominated LSPs mainly depends on the second term in the first line of eq.~(\ref{eqn:ainjnk-reduced-without-approximation}). 
We will utilize these relationships in future discussions to estimate DM relic density and DMDD cross-sections.

The squared mass of the
SM-like Higgs boson, at one-loop level, is given by
\cite{Ellwanger:2011sk,Carena:2015moc}
\beq
\mhsm^2 = m_Z^2 \cos^2 2\beta + \lambda^2 v^2 \sin^2 2\beta + \Delta_\mathrm{mix}+ 
\Delta_{\mathrm{rad.\,corrs.}} \, ,
\label{eqn:hsmmass}
\eeq  
where $\Delta_\mathrm{mix}$ is the contribution from the possible a singlet-doublet mixing and $\Delta_{\mathrm{rad.\,corrs.}}$ is the MSSM-like one-loop radiative  corrections.

The tree-level squared mass of the $CP$-even singlet-like Higgs boson ($h_s$), at the minimal mixing with the doublet-like states, is given by the (3,3) component of ${\cal M}_{S}^2$, i.e.,
\beq
\mhssq \approx {\cal{M}}^2_{S,33} =  
 \lambda \alambda  \frac{v_u v_d}{\vs} + \kappa \vs (\akappa + 4\kappa \vs) \,.
\label{eqn:cp-even-mass}
\eeq

On the other hand, the tree-level singlet-like $CP$-odd Higgs boson squared mass (up to some mixing with its doublet cousin) is given the (2 $\times$ 2) component of the ${\cal M}_P^2$ matrix, i.e.,
\beq
\massq \approx {\cal{M}}^2_{P,22} =  
\lambda (\alambda + 4\kappa \vs)\frac{v_u v_d}{\vs} -3\kappa \akappa \, \vs \, .
\label{eqn:cp-odd-mass}
\eeq
It can be seen from ${\cal M}_{S}^2$ and ${\cal M}_P^2$ that the masses of various Higgs bosons have involved dependency on the six input parameters $\tanb$, $\mueff$, $\lambda$,
$\kappa$, $\alambda$ and $\akappa$. The electroweakino mass matrix ${\cal M}_0$ depends on the first four input parameters with $M_1$ and $M_2$. It is expected that the masses of the singlet-like Higgs bosons ($a_s$ and $h_s$) are connected with the singlino mass ($\msinglino$).

The first term in eq.~(\ref{eqn:cp-odd-mass}) could be ignored at small $\lambda$ and $\massq \approx |-3 \kappa \akappa \vs| = |-{3 \over 2} \akappa \msinglino|$. For a fixed $\msinglino$, the mass of $a_S$ decreases with $\akappa$.
In this study, we focus on the small $\akappa$ region of parameter space for relatively lighter $a_S$.
For larger  $\msinglino$, $\mhs \sim |\msinglino|$ (from eq.~\ref{eqn:cp-even-mass}). For smaller $\msinglino$, $\mhs$ depends on various other parameters and it could be even smaller than $\mhsm$.
In this study, the singlino is relatively light, which corresponds to a relatively lighter $\hs$ in the spectrum. 
The $CP$-even and $CP$-odd doublet-like heavy doublet Higgs boson masses ($m_H$ and $m_A$, respectively) decouple at larger $\alambda$ limit which could be observed from the (1,1) component of ${\cal M}_S^2$ and ${\cal M}_P^2$. Larger $\alambda$ provides the ``alignment via decoupling'' condition in the doublet Higgs sector.
Before ending this section, we would like to highlight the interaction coupling of the $Z$-boson with a pair of neutralinos, as we will refer to it in future discussions.
The same coupling is given by 
\beq
g_{_{Z \ntrlj \ntrlk}} =\frac{g_2}{2 \cos\theta_W}(N_{j3} N_{k3} - N_{j4} N_{k4}) \,,
\label{eqn:znjnk}
\eeq
where $\theta_W$ is the weak mixing (Weinberg) angle.
\subsection{The dark matter sector}
\label{DMsector}
In the presence of $R$-parity-conserving SUSY models, the LSP (R-odd) naturally produces a DM candidate of the universe~\cite{Goldberg:1983nd, Ellis:1983ew}. In this work, we choose singlino-dominated DM with singlino content at least 90$\%$. We consider the Planck reported DM relic density upper bound ($\Omega h^2 \lesssim 0.120$) constraint (to avoid the possibility of an over-closed universe). To meet this constraint, sufficient annihilation of the singlino-dominated LSP in the early universe is needed. The usual resonant $s$-channel annihilation for a pair of LSPs is via the exchange of $Z$-boson, $\hsm$ and the singlet-like Higgs bosons $a_S$, $h_S$~\footnote{Note that constraints arising from searches for heavy doublet-like Higgs bosons ($H$, $A$) at the LHC restrict the possibility of light DM (below 200 GeV) annihilation through their exchange.}.  A pair of DMs may also annihilate through $t$-channel processes via the exchange of a chargino or neutralino. Furthermore, the singlino-dominated DM might co-annihilate with bino-like or higgsino-like NLSP\footnote{As the DM has some bino mixing, DM co-annihilation with sleptons is also possible. But for the present work, this DM annihilation mechanism is not possible as we have fixed all the slepton masses at few TeV.}. It is crucial to emphasize that, for all these processes, the mixing of the singlino-dominated state with the bino-like and the higgsino-like states can play an important role. It can be seen from eq.~(\ref{eqn:znjnk}) and (\ref{eqn:hin1n1-2}) that for the $Z$ and $\hsm$ mediated processes, the singlino-dominated LSP needs at least some higgsino admixtures. This singlino-higgsino mixing significantly depends on  $\lambda$, which controls the strength of some important couplings of those processes. 
It can be observed from eqs. (\ref{eqn:N15-N11relation-1}), (\ref{eqn:N13-N11relation-1}) and (\ref{eqn:N14-N11relation-1}) that at smaller $\lambda$ region a significant bino-mixing (driven by $g_1$ and the soft mass $M_1$) is possible in the singlino-dominated LSP. In this scenario, the $g_1$ proportional second term in eq.~(\ref{eqn:hin1n1-2}), i.e., the gauge interactions with the pair of LSPs, can become significant in $g_{_{\hsm \ntrlone \ntrlone}}$.
Furthermore, it can also be noted from eqs.~(\ref{eqn:N13-N11relation-1}) and (\ref{eqn:N14-N11relation-1}) that such bino mixing could exhibit a non-trivial dependence on the $Z$-boson coupling with a pair of LSPs, as described in eq.~(\ref{eqn:znjnk}) with $j, k =$ 1.
In a subsequent section, we will explicitly demonstrate some intricate dependencies of these couplings on the $M_1$ parameter.
The $a_S$ and $h_S$ mediated DM annihilation processes significantly depend on 
$\kappa$ (via the singlino-singlino-singlet interaction) and $\lambda$ (through the higgsino-higgsino-singlet interaction) parameters.

The null results from numerous searches of DM at various DMDD experiments put strong constraints on both the DM-nucleon SI and SD scattering cross-sections.
In the heavy squark limit, the $t$-channel three $CP$-even scalars ($h_{SM}$, $h_{H}$ and $h_{S}$) mediated processes mostly contribute to the DMDD-SI scattering cross-section. The interactions of these scalars with an LSP pair are given by eq.~(\ref{eqn:hinjnk-reduced-without-approximation}) (with $j=k=0$) and with a pair of nucleons is given by,
\bea
\label{eqn:hiNN-badziak}
g_{_{h_i N N}}
 &= &
\frac{m_N}{\sqrt{2}\, v}\left({S_{i1} \over \cos \beta}  F^{(N)}_d+ {S_{i2}  \over \sin \beta} F^{(N)}_u \right),
\eea   
where $m_N$ is the mass of the nucleon, $F^{(N)}_{d,u}$ are the combinations of various nucleon form factors, refs.~\cite{Belanger:2013oya, Badziak:2015nrb}. Note that this interaction only depends on the doublet admixtures of the $CP$-even Higgs bosons. This implies that a pair of nucleons coupling with the singlet-like $CP$-even Higgs boson is highly suppressed. The SI scattering cross-section $(\sigma^{\rm SI}_{\ntrlone-(N)})$ is given by~\cite{Cao:2018rix}
\beq
\label{eqn:SI-equation-general}
\sigma^{\rm SI}_{\ntrlone-(N)} = \frac{4 \mu^2_r}{\pi} \left|f^{(N)}\right|^2, \hspace{0.5cm} f^{(N)} = \sum_{i=1}^3 {\frac{g_{_{h_{i} \ntrlone \ntrlone}}\, g_{_{h_{i} N N}}}{2 m^2_{_{h_i}}}}  \,, 
\eeq
where $\mu_r$ is the DM-nucleon system reduced mass. We consider the ``$decoupling$ $limit$" condition in the Higgs sector.  In this case, $1/m_H^4$ effectively suppresses the contribution of this decoupled $CP$-even Higgs boson to the SI cross-section.
 
 In the basis of eq.~(\ref{eqn:hiEhhat}), eq.~(\ref{eqn:hiNN-badziak}) is given by~\cite{Badziak:2015exr}
\bea
\label{eqn:hiNN-newbasis}
g_{_{h_i N N}}
 &= &
\frac{m_N}{\sqrt{2} v}  \Big[ E_{h_i \hat{h}} (F^{(N)}_d + F^{(N)}_u) + E_{h_i \hat{H}} \big(\tan\beta F^{(N)}_d - \frac{1}{\tan\beta} F^{(N)}_u\big) \Big] \, .
\eea
The second term in (\ref{eqn:hiNN-newbasis}) suggests that for larger $\tan\beta$, the $g_{_{H N N}}$ coupling is strengthened. In the larger $\tan\beta$ region of parameter space, the mass suppression effect of the heavy doublet-like Higgs, $H$ (with a mass of a few TeV), in the contribution to the DMDD-SI cross-section can be compensated by this enhancement factor.
In such cases, the heavy Higgs boson contribution may become comparable to that of the SM-like Higgs boson at 125 GeV.
The contribution of the singlet-like Higgs boson, $h_S$, is suppressed by its coupling with a pair of nucleons. On the other hand, for the singlino-dominated LSP, the $g_{_{\hs \ntrlone \ntrlone}}$ coupling can become relatively large, resulting in a considerable contribution to the SI cross-section. Additionally, a lighter singlino state correlates to a lighter $\hs$ (even smaller than $\hsm$), which can aid in increasing its contribution to the SI cross-section.  

The region of parameter space where the DMDD-SI scattering cross-section is modest has been pushed by recent experimental limits.  Nonetheless, there are ``blind spot" locations in the parameter space where the neutralino LSP SI scattering cross-section (nearly) vanishes. 
A relatively small $g_{_{\hsm \ntrlone \ntrlone}}$ (the so-called `coupling blind spot') reduces the contribution of the $\hsm$ mediated process to the SI cross-section.
On the other hand, another type of blind spot of the SI cross-section could happen due to the destructive interference among the processes mediated by various $CP$-even scalars.
In the vicinity of
such blind spot conditions, the DMDD-SI cross-section could be brought
down to the irreducible neutrino background (the so-called `neutrino floor/fog' $\sim 10^{-49} \, \mathrm{cm}^2$)~\cite{Billard:2013qya}) of the DMDD experiments. When the squarks are heavy, the $Z$-boson-mediated t-channel process contributes to the DMDD-SD scattering cross-section. A similar type of ``SD blind spot'' could occur when $g_{_{Z \ntrlone \ntrlone}}$ becomes very small.
In the next subsections, we will discuss the conditions of blind spots of SI and SD  scattering cross-sections for the singlino-dominated DM (LSP).
\subsubsection{The SI coupling blind spot}
\label{subsubsec:si}
%%%%%%%%%%%%%%%%%%%%%
The SM-like $CP$-even Higgs, $\hsm$, dominates the DMDD-SI cross-section in the region of parameter space where the other $CP$-even Higgs bosons $H$ and $\hs$ are decoupled and their contributions to the SI scattering cross-section are negligible.
A vanishing $g_{_{\hsm \ntrlone \ntrlone}}$ ($\approx 0$) causes a (coupling) blind spot scenario, which indicates a significant drop in the contributions to the DMDD-SI cross-section via the $\hsm$-mediated process.
The required condition for the coupling blind spot is given by (from eq.~\ref{eqn:hin1n1-3}),
\bea
\label{eqn:coupling-blindspot}
\Bigg[\mntrlone  + {{g_1^2 v^2} \over {M_1 - \mntrlone}} + {g_1^4 v^4  \over 4}{\mntrlone + \mueff \sin 2\beta \over (M_1 - \mntrlone)^2 \, (\mueff^2 - \mntrlone^2)} \Bigg]{1 \over \mueff \sin 2\beta}= 1 \,.
\eea
In the bino decoupling limit ($M_1 >> \mueff, \msinglino$), eq.~(\ref{eqn:coupling-blindspot}) reproduces the ``well-known" coupling blind spot condition for singlino-dominated LSP~\cite{Baum:2017enm} in a singlino-higgsino (3$\times$3) system of neutralinos, i.e.,
\bea
\label{eqn:MSSMcoupling-blindspot}
{\mntrlone \over \mueff} \approx \sin 2\beta \,.
\eea 
This blind spot condition appears only when the $\lambda$ proportional first term in eq.~(\ref{eqn:hin1n1-2}), i.e., the singlino-higgsino-Higgs boson interaction term, is considered.
Note that such a criterion of eq.~(\ref{eqn:MSSMcoupling-blindspot}) can only be satisfied if $\mntrlone$ and $\mueff$ have the same relative sign. Usually, the sign of a singlino-dominated LSP ($\mntrlone$) depends on the sign of $\msinglino$ (2$\kappa\mueff/\lambda$). This implies $\kappa > 0$ to satisfy eq.~(\ref{eqn:MSSMcoupling-blindspot}). It is also worth noting that to meet the aforementioned blind-spot requirement, $\tan\beta$ must grow with $\mueff$ for a fixed $\mntrlone$.

The presence of a relatively light bino-like state (low $M_1$) can open up a new region of parameter space where singlino-dominated DM exhibits a notably low SI scattering cross-section in DMDD experiments.
In comparison to the second term, the third term on the left-hand side of eq.~(\ref{eqn:coupling-blindspot}) is suppressed for $|M_1|-|\mntrlone| \lesssim g_1 v$ and
$|\mueff| - |\mntrlone| \gg g_1 v$. 
In this scenario, it is more probable that $\ntrltwo$ exhibits characteristics of the bino-like state, while the higgsino-like neutralino states correspond to $\ntrlthree$ and $\ntrlfour$.
In this limit, the coupling blind spot condition of eq.~(\ref{eqn:coupling-blindspot}) can be approximated as, 
\bea
\label{eqn:coupling-blindspot-approxi}
\bigg(\mntrlone + {g_1^2 v^2 \over {M_1 - \mntrlone}}\bigg) {1 \over \mueff \sin 2\beta} \simeq 1 \,.
\eea
It is worth noting that the second term on the left-hand side of eq.~(\ref{eqn:coupling-blindspot-approxi}) is derived from the $g_1$-proportional term (i.e., the MSSM origin gaugino-higgsino-Higgs boson interaction term) in eq.~(\ref{eqn:hin1n1-2}).
A further inspection reveals that when $\mntrlone$ and $\mueff$ carry a relatively opposite sign between them, i.e. $\kappa < 0$ region of parameter space,  such agreement of eq.~(\ref{eqn:coupling-blindspot-approxi}) requires $M_1$ and $\mueff$ to have the same relative sign. This opens up a new parameter space for singlino-dominated DM with a significantly small DMDD-SI cross-section. 
eq.~(\ref{eqn:MSSMcoupling-blindspot}) always demands $\kappa > 0$ whereas the criteria of eq.~(\ref{eqn:coupling-blindspot-approxi}) can be fulfilled even in $\kappa < 0$ in the presence of light $M_1$ with same relative sign with $\mueff$.
%Later, we will present benchmark points with $\kappa <0$ and bino NLSP with the same sign with $\mueff$ to have a small DMDD-SI scattering cross-section.
For $\kappa > 0$ situation, i.e. same relative sign between $\mntrlone$ and $\mueff$, condition of eq.~(\ref{eqn:coupling-blindspot-approxi}) can be satisfied irrespective of $M_1$ sign with $\mueff$. In this case, the preferred region for the DM to exhibit a relatively low DMDD-SI scattering cross-section depends on the relative sign and the relative mass splits between the neutralino states and the term $\mueff\sin\beta$.

As previously discussed, in order to satisfy eq.~(\ref{eqn:MSSMcoupling-blindspot}) with $\kappa > 0$, for a fixed mass of the singlino-dominated DM $\tan\beta$ must rise with $\mueff$. 
Thus, for relatively light DM exhibiting a considerably low DMDD-SI cross-section, the parameter space favoring higher values of $\tan\beta$ would also lean toward higher values of $\mueff$.
However, the more generalized blind spot condition expressed in eq.~(\ref{eqn:coupling-blindspot-approxi}) indicates that DMDD-SI rates can become small at relatively low $\tan\beta$ for larger $\mueff$ in the context of relatively light singlino-dominated DM and $\kappa > 0$.
This requires relatively low $M_1$ and prefers relative sign combinations between $\mueff$ and $M_1$. 

The DMDD-SI scattering blind spot conditions due to the destructive interference between the $CP$-even Higgs bosons ($\hsm$, $H$, $\hs$) mediated processes for singlino-dominated DM of a singlino-higgsino (3$\times$3) neutralino system is well studied in the literature~\cite{Badziak:2015exr, Badziak:2015nrb}. However, the presence of a relatively light bino state could alter the blind spot conditions and open up new regions of parameter space of low DMDD-SI scattering cross-section. This requires a more detailed study and we plan to explore it in future work.
\subsubsection{The SD blind spot}
\label{subsubsec:sd}
%%%%%%%%%%%%%%%%%%%%%%%%%%%%%
The $Z$-mediated LSP-nucleon DMDD-SD scattering cross-section $(\sigma^{\rm SD}_{\ntrlone-(N)})$ approximately equals to
\beq
\label{eqn:SD-equation-approx}
\sigma^{\rm SD}_{\ntrlone-(N)} \simeq C_N \times \bigg(\frac{g_{_{Z \ntrlone \ntrlone}}}{0.01}\bigg)^2,
\eeq 
where $C_p \simeq 2.9\times 10^{-41}$ $\text{cm}^2$ for proton and $C_n \simeq 2.3\times 10^{-41}$ $\text{cm}^2$ for neutron. The SD scattering cross-section depends on $g_{_{Z \ntrlone \ntrlone}}$ which is proportional to $N_{13}^2 - N_{14}^2$ (from eq.~\ref{eqn:znjnk} with $j=k=1$)~\cite{Haber:1984rc}. Using eqs. \ref{eqn:N14-N11relation-1} to \ref{eqn:I-value} one can find,
\bea
\label{eqn:sigma-SD-1}
N^2_{13} - N^2_{14}
 =
{\lambda^2 v^2 \over I} \cos 2\beta \Bigg[-1 + {g_1^2 v^2 \over \bigg({M_{_1} \over \mntrlone} - 1 \bigg)(\mueff^2 - \mntrlone^2)} + {g_1^4 v^4  \over 4 (\mone - \mntrlone)^2(\mueff^2 - \mntrlone^2)} \Bigg] \, . \nonumber \\
\eea
At $\tan\beta = 1$, $\cos2\beta$ vanishes, and a well-known MSSM-like DMDD-SD blind spot appears. Another new kind of DMDD-SD blind spot can happen in the NMSSM if the terms inside the bracket of the above expression cancel each other. This can happen most likely when $\mone$ and $\mntrlone \sim \msinglino$ have the same sign. It indicates that the presence of a light bino state could degrade the DMDD-SD scattering cross-section of singlino-dominated DM irrespective of the $\tan\beta$ value. Note that, in the limit $\mntrlone^2 \sim \msinglino^2 \ll \mueff^2, \, \mone^2$,  eq.~(\ref{eqn:SD-equation-approx}) leads to
$\sigma^{\rm SD}_{\ntrlone-(N)} \propto 1/\mueff^4$. Thus, the DMDD-SD scattering cross-section decreases as $\mueff$ increases.
%
%%%%%%%%%%%%%%%%%%%%
\section{The spectra and constraints from various sectors}
\label{constraints}
In our study, we take into consideration a variety of constraints from diverse sectors, spanning both theoretical and experimental domains. The theoretical constraints involve ensuring that the spectra remain free from tachyonic states, preventing the scalar potential from developing an unphysical global minimum, and ensuring that various relevant couplings within the theory do not encounter Landau poles as energy varies, etc. On the experimental front, we incorporate constraints derived from observations in the Higgs sector, DM sector, flavor sectors, and various
new physics searches at the colliders.
We use various publicly available packages like \nmssmtools~{\tt (v5.5.3)}~\cite{Ellwanger:2005dv,Das:2011dg}, 
\higgsbounds~{\tt (v5.8.0)}~\cite{Bechtle:2020pkv}, \higgssignals~{\tt (v2.5.0)}~\cite{Bechtle:2020uwn}, 
\smodels~{\tt (v2.1.3)}~\cite{Alguero:2021dig} and \checkmate~{\tt (v2.0.37)}~\cite{Dercks:2016npn}. 

\nmssmtools ~ is employed to compute the masses, mixings and decays of various 
NMSSM excitations and constrain various relevant observables from the DM, the flavor and the collider sectors.
The above-mentioned theoretical constraints are also checked using \nmssmtools.
Various key observables in the DM sector are obtained using \micromegas{\tt -v4.3}~\cite{Belanger:2006is, Belanger:2008sj, Barducci:2016pcb}, which is integrated within \nmssmtools.
We consider the Planck collaboration~\cite{Planck:2018vyg} measured central value of the DM relic abundance (i.e., $\Omega h^2 = 0.120$) and assume a theoretical uncertainty of 10\% in the estimation of the DM (singlino-dominated LSP) relic density.
The most stringent bounds on the DMDD-SI scattering cross-section are provided by the latest XENON-nT~\cite{XENON:2019rxp, XENON:2023cxc} and PANDA-4T~\cite{PandaX-4T:2021bab} results.
In addition to this, the stringent constraints on DMDD-SD scattering cross-section come from XENON-nT, LUX-ZEPLIN (LZ)~\cite{LZ:2022lsv}, PICO-60~\cite{PICO:2017tgi} and IceCube~\cite{IceCube:2021xzo}.
Indirect searches for DM, conducted through experiments like Fermi-LAT~\cite{Fermi-LAT:2017bpc} and MAGIC~\cite{MAGIC:2016xys}, have led to constraints on the individual thermal average annihilation cross-section of DM into the SM charged pairs, denoted as $\langle \sigma v \rangle_{S S \to \psi \overline{\psi}}$, where $\psi:={\mu, \tau, b, W}$.

Constraints coming from the Higgs sector are checked using
\higgsbounds{\tt -v5.8.0} and
\higgssignals{\tt -v2.5.0}.
Direct searches of various BSM Higgs bosons at LEP, Tevatron and recently at the LHC constraints are examined by the package \higgsbounds.  
On the other hand, one of the $CP$-even Higgs bosons has to 
correspond to the observed Higgs boson at the LHC and coincide its properties with the various measurements of the observed Higgs boson by CMS and ATLAS.
This has been tested using the package \higgssignals.
Note that in our analysis, we consider the mass of the observed SM-like Higgs boson to fall within the range of $122$ GeV $< \mhsm < 128$ GeV. This range accounts for the estimated theoretical uncertainties in its prediction.

The direct searches for electroweakinos, particularly those of relatively light mass, at the LHC are most effectively conducted through the examination of final states characterized by multiple leptons and jets along with large missing energy. 
These states arise from the direct production of $\charonepm \ntrltwo$, with the assumption that both the charged $\charonepm$ and neutral $\ntrltwo$ particles are degenerate and wino-like, while the LSP is bino-like. These events subsequently undergo cascades, manifesting in two distinct decay modes: $WZ\ntrlone$ and $W\hsm\ntrlone$. In the former mode, the final state is characterized by $3\ell + \etmiss$, with both $W^\pm$ and $Z$ bosons decaying leptonically~\cite{CMS:2017moi, CMS:2018szt, ATLAS:2018ojr, ATLAS:2018eui, 
ATLAS:2019wgx, ATLAS:2020ckz, CMS:2021edw}.
Alternatively, the cascade may result in the $2\ell + 2$-${\rm jet} + \etmiss$ final state, where $W^\pm$ bosons decay hadronically~\cite{CMS:2017moi, CMS:2018szt, ATLAS:2018ojr, ATLAS:2018eui}.
Additionally, in the latter mode, another type of final state  characterized by $1\ell + 2b$-${\rm jet} + \etmiss$ can occur, where the $b$-jets originate from the decay of the Higgs boson ($\hsm$)~\cite{CMS:2017moi, CMS:2018szt, ATLAS:2018qmw, CMS:2019pov, 
ATLAS:2020pgy, ATLAS:2020ckz}.
This scenario has further been studied considering the di-photon decay mode of $\hsm$~\cite{ATLAS:2020qlk}.
It is also possible to search these electroweakinos from the pair production of the charginos, i.e., $\charonepm \charonemp$, in the final state of $2\ell + \etmiss$~\cite{ATLAS:2019lff,CMS:2018xqw}.
These investigations over the years represent a critical aspect of the efforts to explore electroweakinos and their properties at the LHC. However, null results from various direct
searches of electroweakinos at the LHC in these modes put constraints on their masses.
%Note that, \nmssmtools~ takes into consideration, albeit in a simplified manner, the constraints arising from the CMS analysis of electroweakinos searches in the $3\ell + \etmiss$ final state, based on 35.9~\fbinv~of data~\cite{CMS:2018szt}.

In the scenario where the concerned chargino-neutralinos are higgsino-like (wino-like states are considered to be very heavy), the corresponding mass-bounds are expected to be relaxed given their smaller combined direct production cross-section at the LHC.
On top of that, within certain regions of the parameter space, the heavier neutralinos exhibit suppressed branching ratios to decay into $Z$ and $\hsm$, which can further erode these experimental constraints. Recently it has been shown within the framework of the $Z_3$-symmetric NMSSM that the potential cascade decays of these particles, involving new singlet-like scalars, singlino-dominated and bino-like neutralinos, have the capacity to significantly reduce the sensitivities of current searches at the LHC~\cite{Abdallah:2019znp, Abdallah:2020yag, Chatterjee:2022pxf, Datta:2022bvg}.

The mass bounds derived from the LHC collaborations exhibit a significant reduction in the compressed region, wherein the mass difference between the wino-like neutralino/chargino and the bino-like LSP falls below the mass of the $Z$ boson. In this regime, the cascade decays involve off-shell $Z$, $\hsm$, and $W$ bosons, leading to subsequent decays yielding soft leptons, jets, missing energy, etc~\cite{ATLAS:2019lng, ATLAS:2018ojr,ATLAS:2019lff,CMS:2020bfa,ATLAS:2021moa,CMS:2021cox,CMS:2021few,ATLAS:2021yqv,ATLAS:2022zwa,CMS:2022sfi, CMS:2019san}. This drop in the mass bounds typically persists as the mass splitting decreases, resulting in even softer final-state leptons and jets.
In this study, our focused investigation is aimed at specific regions within the parameter space, primarily motivated by DMDD blind spots. Notably, these regions may also emerge in the context of the compressed regime. Within this framework, we identify regions where the NLSP, whether bino-like or higgsino-like, exhibits significant decay branching fraction to a photon and the singlino-dominated LSP ($\ntrltwo \rightarrow \ntrlone \gamma$).
Additionally, we uncover a novel parameter space region where both higgsino-like neutralino states ($\ntrltwo$ and $\ntrlthree$) exhibit considerable radiative decay rates, characterized by larger branching fractions for the $\ntrltwo \rightarrow \ntrlone \gamma$ and $\ntrlthree \rightarrow \ntrltwo \gamma$ decay modes. These distinctive decay patterns give rise to potentially intriguing collider signatures at the LHC. 
Note that we do not focus on the non-compressed scenario at the LHC around the newly identified DMDD blind spot regions in this work, as it would lead to scenarios similar to those studied in refs~\cite{Abdallah:2019znp, Abdallah:2020yag, Chatterjee:2022pxf, Datta:2022bvg}~\footnote{These papers extensively discuss various possible collider signals involving singlino, bino, and higgsino-like states in the presence of singlet-like light scalars. The newly identified DMDD-SI blind spot region in the non-compressed scenario would also imply similar collider signals, as studied in those works.}. 

It is noteworthy that some of the channels mentioned above in the compressed region have not been explored in previous LHC searches, leading to a significant relaxation of the mass bounds discussed earlier. In particular, such relaxations for the higgsino-like electroweakinos could significantly extend the parameter space, allowing lower values of $|\mueff|$, and can enhance the `naturalness' of the scenario to a considerable degree~\cite{Barbieri:1987fn, Ellis:1986yg, Giudice:2013yca}.
We delve into a detailed discussion of these decay channels in the subsequent sections. 
In order to assess the viability of our benchmark scenarios in passing all pertinent LHC analyses, we utilize \smodels~ and \checkmate~ packages for the recasting of numerous relevant LHC analyses, including recent ones conducted with 139 fb$^{-1}$ of data.
 Both the recast packages calculate a 
`$r$'-value, where $r = (S-1.64 \Delta S)/ S95$,
with `$S$', $S95$ and $\Delta S$ indicating the predicted number of signal events and the experimental limit on `$S$' at 
95\% confidence level and the associated Monte Carlo error, respectively. Nominally, $r > (<) 1$ denotes the 
benchmark point to be disallowed (allowed).  To take into account the significant NLO+NLL 
contributions, all the production cross-sections of the electroweakinos have been multiplied by a
$k$-factor of 1.25~\cite{Fiaschi:2018hgm}.
\section{The motivated region of parameter space}
\label{subsec:motivatedregion}
The choice of the region within the $Z_3$-symmetric NMSSM parameter space for our present work is influenced by several considerations that pertain to both the phenomenology observed at the LHC and its relevance to the DM sector.
We concentrate on the singlino-dominated DM (LSP), which corresponds to the parameter space region with $|\kappa| < \lambda/2$. This prompts our search for the lower $\kappa$ region. As demonstrated in section~\ref{DMsector}, bino tempering within singlino-dominated DM can uncover a novel parameter space region where the DMDD scattering cross-section significantly decreases. We specifically investigate a moderately low $\lambda~(\lesssim 0.2)$ region, where this tempering can become substantial.

In the Higgs sector, our particular focus lies on the relatively light singlet-like Higgs bosons and the observed SM-like doublet Higgs boson.  The MSSM-like Higgs bosons ($A$, $H$, $H^{\pm}$) are considered to be significantly more massive and effectively decoupled from the rest of the system. This decoupling is achieved by selecting a large value for $\alambda$.
Lighter singlet-like scalars can act as a funnel for the moderately tempered light DM mutual annihilation when $\mas$ or $\mhs \sim 2\mntrlone$.
A lower singlino mass ($\msinglino = 2\kappa \mueff/\lambda$) results a relatively lighter $\hs$ (with $\mhssq \sim \msinglinosq$ as indicated in eq.~(\ref{eqn:cp-even-mass})). For a relatively lighter $\as$, it is favorable to have smaller $\akappa$ (with $\massq \sim \akappa \msinglino$ as described in equation~(\ref{eqn:cp-odd-mass})).
In the neutralino sector, a moderate singlino-bino mixing necessitates a lighter bino state, which motivates to scan over a moderate range of $\mone$. 
The latest constraints from LHC and DMDD experiments put the lighter $\mueff$ region of parameter space under tension. 
This encourages us to scan $\mueff$ from a smaller to a larger value.
For a relatively low $\mntrlone$ and moderately large $\mueff$ (i.e., for low $\mntrlone/\mueff$)
 a larger value of $\tanb$ is required for the blind
spot to work (eq.~(\ref{eqn:MSSMcoupling-blindspot})), leading to an acceptably small DMDD-SI rates. 
For smaller $\mueff$ (i.e., relatively larger $\mntrlone/\mueff$), a larger $\sin2\beta$ value eventually requires and hence a smaller value of $\tan\beta$ to satisfy the blind spot criteria of eq.~(\ref{eqn:MSSMcoupling-blindspot}).
This encourages us to look at the wide range of $\tan\beta$.

The Higgs sector of the NMSSM experiences additional tree-level contributions to $\mhsm$ (as shown in eq.~(\ref{eqn:hsmmass})) compared to those in the MSSM~\cite{Ellwanger:2009dp}.
This extra tree-level contribution becomes substantial at smaller $\tan\beta$ and larger $\lambda$ regions.
In such regions, it is possible to attain the mass of the observed $\hsm$ ($\mhsm \sim$ 125 GeV) without the need for substantial radiative corrections coming from the top squarks.
Conversely, in regions with relatively lower $\lambda$ and higher $\tan\beta$, substantial radiative corrections are necessary to achieve $\mhsm \sim$ 125~GeV.
This motivates us to explore a broad range of larger values for $A_{\text{top}}$ and the masses of third-generation squarks ($m_{Q_3}$ and $m_{U_3}$). Table \ref{scanranges} presents the scan ranges considered for various model parameters. All input parameters of  \nmssmtools~are defined at the scale $Q^2 = (2m_{\tilde{Q}}^2 + m_{\tilde{U}}^2+m_{\tilde{D}}^2)/4$, except for $\tanb$, which is defined at $m_Z$~\cite{Ellwanger:2005dv}.
\begin{table}[t]
\begin{center}
\begin{tabular}{|c|c|c|c|c|c|c|c|c|c|}
\hline
$\lambda$ & $|\kappa|$ & $\tanb$& \makecell{$|\mueff|$ \\ (GeV)}&  \makecell{$|\alambda|$ \\ (TeV)} &
\makecell{$|\akappa|$ \\ (GeV)}
 & \makecell{$|\mone|$ \\ (GeV)} & \makecell{$|A_{top}|$ \\ (TeV)} & \makecell{$M_{Q_{3}}$ \\ (TeV)} & \makecell{$M_{U_{3}}$ \\ (TeV)} \\
\hline
0.001--0.2 &$\leq 0.1$&
1--60& $\leq 750$ & $\leq 5$ & $\leq 300$ & $\leq$600& $\leq 5$& 2--5& 2--5\\
\hline
\end{tabular}
\caption{Ranges of various NMSSM   parameters adopted for
scanning the parameter space.}
\label{scanranges}
\end{center}
\end{table} 
%
%%%%%%%%%%%%%%%%%%%%%%%%
\section{Probing DM with photons at the LHC}
%%%%%%%%%%%%%%%%%%%%%%%%
\label{DMsearchwithphoton}
In this section, we focus on the investigation of various search channels of DM characterized by final states comprising photons at the LHC. These channels are contingent upon the intricate cascade decays of the heavier electroweakino states and their interactions with the DM.
One particularly effective mechanism for DM annihilation involves co-annihilation with heavier sleptons, neutralinos and/or charginos~\cite{Ellis:1998kh,Ellis:1999mm,Buckley:2013sca,Han:2013gba,Cabrera:2016wwr,Baker:2018uox}.
The significance of these annihilation processes of DM depends on the condition that the cross-section for processes involving the heavier co-annihilating particle ($X_2$), with a mass denoted as $m_{X_2}$, significantly exceeds that of the pair of DM ($\ntrlone$) annihilation processes.
If the mass difference between $X_2$ and $\ntrlone$ remains relatively modest, allowing $X_2$ to persist within the thermal plasma during DM freeze-out, co-annihilation processes can efficiently reduce the relic density of the DM, despite the Boltzmann-suppression factor ($\exp\left[ - \left( m_{X_2}-\mntrlone \right) / T \right]$).
This factor accounts for the difference in number densities between $X_2$ and $\ntrlone$. In addition to this, ``assisted co-annihilation'' can significantly enhance the DM annihilation cross-section~\cite{Ellwanger:2009dp, Ellwanger:2016sur}, particularly when the annihilation cross-section of the pair of $X_2$ particles into the SM particles is notably large. 
For example, this is often the case when $m_{X_2}$ closely approaches half of the mass of one of the Higgs bosons, giving rise to a resonant annihilation process of a pair of $X_2$ particles that can overcome the Boltzmann-suppression factor.
Nevertheless, a critical hallmark of the co-annihilation mechanism is the presence of a small mass gap between the heavier particle and the DM. This feature prevents the Boltzmann suppression factor from reaching excessively high values and makes the scenario significantly compressed, i.e., $(m_{X_2} -\mntrlone) < m_Z$.

In our scenario, singlino-dominated DM (LSP) can co-annihilate with the bino-like neutralino or the higgsino-like neutralinos/charginos.
Various decay branching fractions of the heavier states, like the bino-like neutralino or the higgsino-like neutralinos/charginos, to the LSP depend on their couplings and the available phase space. 
In the compressed regions, all the decay modes of the second neutralino, $\ntrltwo$, are kinematically suppressed, and it can be parametrized by the well-known “mass splitting parameter"~\cite{Baum:2023inl} 
\begin{equation}
    \varepsilon \equiv \frac{\mntrltwo}{\mntrlone}-1 \, \,.
\end{equation}
It is interesting to note that tree-level decay processes of $\ntrltwo$, such as $\ntrltwo \rightarrow \ntrlone f \bar{f}$ mediated by off-shell $Z$, $\hsm, H, A, h_s, a_s$, experience a suppression of $\varepsilon^5$, whereas radiative decay processes, such as $\ntrltwo \rightarrow \ntrlone \gamma$ which are induced by triangle loops involving fermions, sfermions, charginos, neutralinos, charged Higgs bosons, and $W^{\pm}$, are subject to a suppression of $\varepsilon^3$~\cite{Ambrosanio:1996gz}. Therefore, in the compressed region, radiative decay of the heavier neutralino states can play a pivotal role. 
The radiative decay branching fraction of the heavier neutralinos, $\chi_{i}^0$, can experience a further significant enhancement when the couplings $g_{\chi_{i}^0 \ntrlone Z}$ and $g_{\chi_{i}^0 \ntrlone \hsm}$ become significantly small and/or the effective coupling $g_{\chi_{i}^0 \ntrlone \gamma}$ increases.
This scenario corresponds to specific regions within the parameter space where such conditions are met.

It has been demonstrated that the couplings $g_{\chi_{i}^0 \ntrlone Z}$ and $g_{\chi_{i}^0 \ntrlone \hsm}$ in the singlino-higgsino neutralino sector crucially depend on the value of $\lambda$~\cite{Abdallah:2020yag}. The coupling
$\lambda$ not only drives the higgsino-singlino-Higgs boson (scalar) interaction (see (eq.~\ref{eqn:hinjnk1})) involved in the decay $\chi_{i}^0 (\tilde{H}) \rightarrow \hsm \chi_1^0 (\tilde{S})$, but regulates the mixing in the higgsino-singlino sector, exerting a crucial influence on the associated couplings.
Conversely, the dependence on $\lambda$ in the coupling of the $Z$-boson to the neutralino states solely appears through mixing. 
 It has been shown explicitly in reference~\cite{Abdallah:2020yag} that these couplings increase with $\lambda$. Thus, the radiative decay branching fraction of the higgsino-like neutralinos can increase via decreasing $g_{\chi_{i}^0, \ntrlone Z}$, $g_{\chi_{i}^0, \ntrlone \hsm}$ couplings at the lower $\lambda$ region of parameter space.
Furthermore, it is intriguing to note that the presence of a light bino-like state ($\ntrlfour$) close to the higgsino-like and singlino-dominated states can increase the radiative decay branching fraction of the higgsino-like states ($\ntrltwothree$).
In addition to this, a light bino state in the proximity of the higgsino and singlino-dominated states can reduce the DMDD scattering cross-section, as we discussed in section~\ref{DMsector}.
 Thus, bino can play an important role in the search for DM in both collider and DMDD experiments.

Turning to the relevant decays that shape the phenomenology at the LHC, one of the interesting channels to search this singlino-higgsino co-annihilation region of parameter space from the production of $pp \rightarrow \ntrltwothree (\tilde{H}) \charonepm (\tilde{H})$ could be $1\ell + \geq 1\gamma + \etmiss$ where leptonic decay of $W^{\pm}$ is considered and the higgsino-like $\ntrltwothree$ branching ratio to radiative decay mode is significantly large. 
Due to the small mass gap between higgsino-like states and the singlino-dominated LSP, it is expected that lepton and photons will be relatively soft in this channel.
Note that, in reference~\cite{ATLAS:2020qlk}, ATLAS has already conducted a search for the neutralino/charginos, considering a similar final state at the LHC.
However, in this analysis, it is considered that the pair of photons originate from the decay of the on-shell $\hsm$, which arises from the decay of heavier neutralino.
Therefore, this analysis does not remain relevant to the region of parameter space of the singlino-higgsino co-annihilation that we focus on in this work. 
On the other hand, in scenarios where the decaying leptons/jets are too soft to be detected, the production of higgsino-like states together with an ISR jet or a photon can become relevant for the mono-jet/$\gamma$ + $\etmiss$ LHC analyses~\cite{ATLAS:2018nud, ATLAS:2017bfj, ATLAS:2021kxv,CMS:2021far}.
 This can impose reasonable constraints on the chargino-neutralino sector in this compressed scenario. However, mono-jet/mono-photon searches at the LHC remain insensitive to the production of a pair of singlino-dominated LSPs due to its inherently small production cross-section.

In the case of singlino-bino co-annihilation, NLSP is bino-like, and it could also have a significant radiative branching fraction.
Therefore, in this scenario, cascades of the produced higgsino-like $\ntrlthreefour$ and $\charonepm$ via $\ntrltwo$ could lead to: $pp \rightarrow \ntrlthreefour (\widetilde{H}) \charonepm (\widetilde{H}) 
\rightarrow $ $\hsm/Z + W^{\pm} + \ntrltwo (\tilde{B}) \, \big[\ntrltwo \rightarrow  \gamma \, \ntrlone (\tilde{S})\big]$
%$\rightarrow $ $\hsm/Z + W^{\pm} + \etmiss$ + $\geq  1\gamma$
$\Rightarrow 3\ell + \geq 1\gamma + \etmiss$ or $1\ell + 2b + \geq 1\gamma + \etmiss$ final states at the LHC.
Similar to the previous scenario, due to a small mass gap, photons are also anticipated to remain soft in this context. But, for a larger mass gap between the higgsino-like states and the bino-like NLSP, the NLSP can emerge with a substantial boost resulting from the decay of those higgsino-like states. In such cases, the photons originating from the NLSP decay no longer remain soft.
On the other hand, the leptons and jets coming from the on-shell $W^{\pm}, \hsm, Z$ bosons are expected to be relatively hard.
In some cases, if the photons are soft and remain undetected, cascades of $\ntrlthreefour$ and $\charonepm$ via bino-like $\ntrltwo$ would lead to similar final states as when these higgsino-like heavier neutralino states directly decay to the singlino-dominated LSP.
Thus, virtually, a large effective branching ratio to the LSP for
these states could be envisaged, and stringent experimental lower bounds, which we discuss in section~\ref{constraints}, will be applicable to their masses. 
Note that, as pointed out earlier, since the ATLAS analysis~\cite{ATLAS:2020qlk} is dedicated to probing the neutralinos/charginos by examining the di-photon decay mode of $\hsm$ resulting from the decay of the heavier neutralino, certain signal regions involving photons can become sensitive to the singlino-bino co-annihilation scenario.
We will delve into this in detail in Section~\ref{benchmark}.

%
%%%%%%%%%%%%%%%%%%
%
\begin{figure}[t!]
\begin{center}
\includegraphics[height=5.9cm,width=0.49\linewidth]{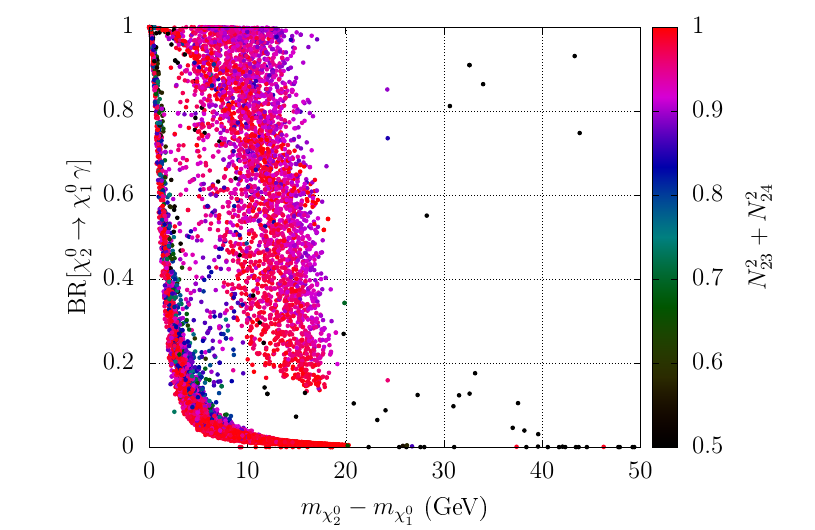}
\hskip 5pt
\includegraphics[height=5.9cm,width=0.49\linewidth]{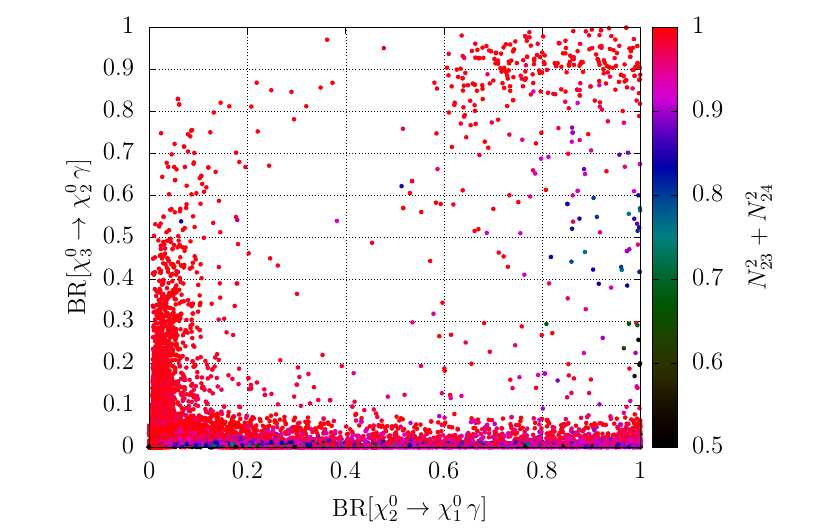}
\caption{Scattered points showing the variation
of BR[$\ntrltwo \rightarrow \ntrlone \gamma$] with $\mntrltwo - \mntrlone$ (left plot) and BR[$\ntrlthree \rightarrow \gamma \ntrltwo$] with BR[$\ntrltwo \rightarrow  \ntrlone  \gamma$] (right plot).
The higgsino content of the NLSP ($N_{23}^2+N_{24}^2$) is shown through the palette in both plots.
The color palettes in these two plots are truncated at the higgsino content of 0.5 for enhanced color clarity, although these plots include points with values below that threshold, i.e., points with $N_{23}^2 + N_{24}^2 < 0.5$ are represented by black points.}
\label{fig:bp2_3photondecay}
\end{center}
\vspace{-0.5cm}
\end{figure}
%%%%%%%%%%%%
%
Since the radiative decays of the heavier electroweakinos are the focus of our present study, we illustrate the variation of the decay branching fraction of $\ntrltwo \rightarrow \gamma \ntrlone$ process with the mass difference between $\ntrltwo$ and $\ntrlone$ in the left plot of figure~\ref{fig:bp2_3photondecay}. Here, LSP is singlino-dominated, i.e., $N_{15}^2 > 0.9$. 
The total higgsino component of the NLSP ($N_{23}^2+N_{24}^2$) is presented through the palette.
Points ranging from red to blue denote highly higgsino-like NLSP, while black points represent mostly bino-like NLSP (i.e., $N_{21}^2 > 0.5$ and $N_{23}^2+N_{24}^2 < 0.5$).
It is noteworthy that, as expected, the BR[$\ntrltwo \rightarrow \gamma \ntrlone$] increases with the decrease of $\mntrltwo - \mntrlone$.
The figure reveals two distinct band patterns. 
For BR$[\ntrltwo \rightarrow \gamma \ntrlone] > 0.9$, the mass gap needs to be below 3 GeV in the lower band. In contrast, in the upper band, this can be achievable even with a mass gap exceeding 15 GeV.
It is observed that the upper band is only possible for $\mueff <0$.
It can be observed that in this upper band, for a fixed BR, the mass gap increases as the higgsino component in the NLSP decreases.
This happens due to the presence of a nearby bino-like state (as $\ntrlfour$), which modifies the mixing in the higgsino-like NLSP in such a way that the radiative decay involved effective coupling increases significantly, and because of that, the BR remains large even at the relatively larger mass difference.

\begin{table}[t]
\renewcommand{\arraystretch}{1.6}
\begin{center}
\begin{tabular}{|c|c|c|c|c|}
\hline
$\kappa$ & $\mueff$ & $M_1$ \\
\hline
\multirow{2}{*}{$-$} & $+$  & $+$ \\
\cline{2-3}
 &  $-$  & $-$ \\
\hline
\multirow{2}{*}{$+$} &$+$  & $-$ \\
\cline{2-3}
 &   $-$  & $+$ \\
\hline
\end{tabular}
\caption{Preferred sign combinations among $\kappa$, $\mueff$, and $M_1$ based on scan results.}
\label{tab:signcomb}
\end{center}
\end{table} 

The decay patterns of the heavier electroweakinos are also crucial in the context of their searches at the LHC. In the right plot of Figure~\ref{fig:bp2_3photondecay}, we present the variation of BR[$\ntrltwo \rightarrow \gamma \ntrlone$] with BR[$\ntrlthree \rightarrow \gamma \ntrltwo$] based on our scan results. Once again, the LSP is singlino-dominated. The higgsino component of the NLSP is displayed in the palette.
It is noteworthy that in our scenario, when $\ntrltwo$ is higgsino-like, the nearly degenerate $\ntrlthree$ will also be higgsino-like, and $\ntrlfour$ will be bino-like.
In this plot, an intriguing observation is the existence of a region in parameter space where both BR[$\ntrltwo \rightarrow \gamma \ntrlone$] and BR[$\ntrlthree \rightarrow \gamma \ntrltwo$] are significantly large. This scenario corresponds to the singlino-higgsino co-annihilation.
The existing multi-lepton $+ \etmiss$ searches for compressed higgsino-like neutralinos at the LHC become less sensitive in this region of parameter space.
In this scenario, we also find that the higgsino-like $\charonepm$ state tends to prefer decaying to an off-shell $W$ boson and $\ntrltwo$ rather than $\ntrlone$.
The significant radiative decay branching fractions of higgsino-like electroweakinos underscore the need for a crucial new detection channel at the LHC involving triggered photons, essential for probing this compressed singlino-higgsino scenario in the NMSSM parameter space.

To illustrate our scenario, in the next subsection, we will present a few benchmark points (BP1 to BP4) featuring larger radiative decay branching fractions of heavier electroweakinos and pass all the existing bounds that are discussed in section~\ref{constraints}.
We also present one benchmark scenario (BP-D1), which is excluded by the recent ATLAS analysis, to demonstrate some interesting collider aspects. 
Before delving into the discussion of those benchmark points, here we will explore the implications of the favored sign combinations among $\kappa$, $\mueff$, and $M_1$.
%, as uncovered by our scan results. 
These combinations shed light on the regions of the parameter space that are more favored, particularly in the context of the latest experimental constraints.
The parameter space encompassing the sign combinations listed in Table~\ref{tab:signcomb} constitutes the majority of the allowed points based on our comprehensive scan results.
 When $\kappa < 0 $, both $\mueff$ and $M_1$ tend to share the same relative sign, while for $\kappa > 0$ they show a preference for opposite relative signs.
It is noteworthy that this favoured pattern can be comprehended through the unique relationship that we introduce for the first time in this work in eq.~(\ref{eqn:coupling-blindspot-approxi}). Such sign combinations are required for the blind spot to work for the singlino-dominated DM, leading to an acceptably small DMDD-SI scattering cross-section.

Previous studies on the DMDD-SI blind spot criterion for singlino-dominated DM only discussed the same-sign condition between $\mntrlone$ and $\mueff$, i.e., $\kappa > 0$ (eq.~(\ref{eqn:MSSMcoupling-blindspot})). In contrast, in this work, we introduce a more general blind spot criterion (eq.~(\ref{eqn:coupling-blindspot-approxi})) which points out that for $\kappa > 0$, $\mueff$ and $M_1$ should have opposite relative signs. Furthermore, this generalized condition reveals a new blind spot region, i.e.,  $\kappa < 0$, which was not previously addressed in the literature.
 It is worth noting that this new region may have significant implications for explaining the discrepancy of the anomalous Muon magnetic moment ($a_{\mu} = (g_{\mu}-2)/2$) between the experimental observations (from Fermilab and BNL) and the SM prediction~\cite{Muong-2:2006rrc, Muong-2:2021ojo}.
The measured value of $a_{\mu}$ is larger than the predictions from the SM. Therefore, a positive contribution from the new physics is required to explain this discrepancy.
In this scenario, it is well-known that the most important
contributions arise from a Bino-smuon and a chargino-(muon-sneutrino) loop. 
Interestingly, the sign of these loop contributions depends on the sign of $(M_1 \times \mueff)$ and $(M_2 \times \mueff)$, respectively. They contribute positively when their signs are positive.
Therefore, we find that in the case of singlino-dominated DM, the region with $\kappa < 0$ prefers the DMDD-SI blind spot scenario if $M_1$ and $\mueff$ have the same relative sign, which also provides a positive contribution from the Bino-smuon loop to $a_{\mu}$. 
This indicates an intriguing correlation in explaining the observed discrepancy of the anomalous muon magnetic moment and the DMDD-SI blind spot conditions.
However, as emphasized before, in this work, we do not study this issue as we consider smuons to be heavy.
 We reserve this study for future work.
\subsection{Benchmark scenarios}
\label{benchmark}
Three broad classes of benchmark scenarios, BP-D1 and BP1 to BP4, are presented in table~\ref{tab:benchmarks} based on the mode of annihilation of the highly singlino-dominated DM.
These include (i) singlino-bino co-annihilation (BP-D1 and BP1), (ii) the $\hsm$ funnel (BP2) and (iii) singlino-higgsino co-annihilation (BP3 nad BP4). Note that all the benchmark points feature photons from the cascades of the heavier electroweakinos to the DM. 

In BP-D1, an efficient singlino-dominated DM annihilation with DM mass $\sim 55$~GeV is achieved via co-annihilation with a bino-like NLSP with mass~$\sim 66$~GeV. The resonant s-channel process of $\ntrlone \ntrltwo \rightarrow \hsm$ provides the most efficient DM annihilation process to comply with the Planck-observed upper bound on the DM relic abundance.
 As described in section~\ref{DMsector}, the tempering of the singlino-dominated DM by a light bino-like state provides new mechanisms
in moderating $g_{\hsm \ntrlone \ntrlone}$ and $g_{Z \ntrlone \ntrlone}$ couplings in such ways that the DMDD-SI and DMDD-SD rates, respectively, could be tamed to comply with their experimentally allowed values.
In addition to this, because of the small mass gap between $\ntrltwo$ and $\ntrlone$ ($\mntrltwo - \mntrlone \sim 10$~GeV), the bino-like NLSP decays almost 100\% to the LSP with a photon.
As $\lambda <  0.1$, BR[$\tilde{H} \rightarrow \tilde{S} \, \, W/Z/\hsm$] is much smaller than the BR[$\tilde{H} \rightarrow \tilde{B} \, \, W/Z/\hsm$]. 
This is since interactions of $\tilde{S}$-$\tilde{H}$-$Z/\hsm/W$ decreases with the decrease of $\lambda$~\cite{Abdallah:2020yag}.
Therefore, the photon count is enhanced from the cascades of the higgsino-like states, as they prefer to decay to the bino-like NLSP, and subsequently, the NLSP undergoes a radiative photon decay.
If photons are sufficiently energetic to be detected, this could signify a substantial increase in yields from the production process of $pp \rightarrow \ntrlthreefour \charonepm$ in the final states $3\ell + \etmiss$ and $1\ell + 2b + \etmiss$,  when photons are incorporated, as opposed to states without the inclusion of photons.

On the collider front, a dedicated \checkmate~analysis rules out BP-D1 (with $r$=$2.87$) via an ATLAS 
analysis~\cite{ATLAS:2020qlk} of 139~fb$^{-1}$ worth data for the search of chargino-neutralinos by studying the di-photon decay channel of the on-shell $\hsm$, which results from the decay of a heavier neutralino.
It is not primarily motivated to explore this specific co-annihilation scenario. However, some signal regions overlap, as the analysis encompasses final states involving leptons, jets, and photons combined with missing energy. 
This analysis focuses on the significant mass gap between the heavier neutralino/chargino and the LSP regions, ensuring that the decaying objects (jets, leptons, photons) maintain sufficiently larger $P_T$.
In this benchmark point, due to a larger mass gap ($\sim 350$~GeV) between $\mueff$ and $M_1$, bino-like NLSP emerges with a boost. This induces some boost in the photon originating from the decay of the NLSP.
As a result, the tail of the invariant mass of the two photons from the process $pp \rightarrow \ntrlthreefour \charonepm$ broadens relatively and lies around the mass window of $\hsm$, which is considered in the selection cuts of this ATLAS analysis.
The signal regions, particularly `Category-12' ($N_j \geq 2$) and `Category-4' ($N_l \geq 1$), gain sensitivity and begin to exclude specific regions in the parameter space of the singlino-bino co-annihilation scenario as they inclusively consider leptons/jets in final states of their analyses. To demonstrate this collider aspect of this scenario, we retain this point as a benchmark.

As emphasised above, we present another singlino-bino co-annihilation benchmark point, BP1, featuring similar characteristics of BP-D1, except $|\mueff|$ is now on the higher side $\sim 700$~GeV.
%,  avoiding these experimental constraints.
Larger $\mueff$ reduces the yields by reducing the cross-section of the process $pp \rightarrow \ntrlthreefour \charonepm$ at the LHC, consequently lowering the sensitivity of the aforementioned ATLAS analysis~\cite{ATLAS:2020qlk}.
A dedicated \checkmate~analysis, taking into account all available analyses in its repository, indicates that BP1 is allowed with $r = 0.68$.
Note that, similar to the previous benchmark point
as $\lambda$ is small, BR[$\tilde{H} \rightarrow \tilde{S} \, \, W/Z/\hsm$] is much smaller than the BR[$\tilde{H} \rightarrow \tilde{B} \, \, W/Z/\hsm$]. 
Furthermore, here BR$[\ntrltwo \rightarrow \ntrlone \gamma]$  is around 88\%.
Therefore, similar to BP-D1, since the higgsino-like states prefer to decay mostly to the NLSP, there is a significant suppression of the yields from the production process of $pp \rightarrow \ntrlthreefour \charonepm$ in the canonical final states $3\ell + \etmiss$ and $1\ell + 2b  + \etmiss$ with no involvement of photons, which are usually considered in the LHC analyses. On the other hand, as we mentioned earlier, if some of the photons that originated from $\ntrltwo$ remain soft to be detected, the cascades of $\charonepm$ and $\ntrlthreefour$ via bino-like $\ntrltwo$ lead to similar final states as when these heavier higgsino-like electroweakinos directly decay to the singlino-dominated LSP.
In this scenario, leptons and jets originating from on-shell $W, Z, \hsm$ are relatively hard.
Therefore, observing excess over the SM backgrounds in both the signals with and without soft photons with  $3\ell + \etmiss$ and $1\ell + 2b  + \etmiss$ at the HL-LHC may point back to the scenario that includes singlino-bino co-annihilation with relatively heavier higgsinos. 
In the next section, we will discuss this by presenting distributions of some kinematic variables of this scenario.

%%%%%%Final table%%%%%%%%%%%%
%
%%%%%%%%%%%%%%%%
\begin{table}[h!]
\renewcommand{\arraystretch}{1.3}
%\begin{center}
\centering{
{\tiny\fontsize{5.5}{5.6}\selectfont{
\begin{tabular}{|c|@{\hspace{0.06cm}}c@{\hspace{0.06cm}}|@{\hspace{0.06cm}}c@{\hspace{0.06cm}}|@{\hspace{0.06cm}}c@{\hspace{0.06cm}}|@{\hspace{0.06cm}}c@{\hspace{0.06cm}}|@{\hspace{0.06cm}}c@{\hspace{0.06cm}}|}
\hline
\Tstrut
\scalebox{1.2}{Input/Observables}  & \makecell{\scalebox{1.2}{BP-D1}} & \makecell{\scalebox{1.2}{BP1}} & \makecell{\scalebox{1.2}{BP2}} & \makecell{\scalebox{1.2}{BP3}} & \makecell{\scalebox{1.2}{BP4}}\\
\hline
\Tstrut
\scalebox{1.3}{$\lambda$}  &  \scalebox{1.1}{$0.0964$} &  \scalebox{1.1}{$0.0964$} &  \scalebox{1.1}{$0.2086$}  &  \scalebox{1.1}{$0.067$} &  \scalebox{1.1}{$0.018$}  \\
\scalebox{1.3}{$\kappa$}   &  \scalebox{1.1}{$0.0062$} &  \scalebox{1.1}{$0.0038$} &  \scalebox{1.1}{$0.0118$} & \scalebox{1.1}{$0.0316$} &  \scalebox{1.1}{$-0.0083$}    \\
\scalebox{1.3}{$\tan\beta$}      & \scalebox{1.1}{10.06}  &  \scalebox{1.1}{7.01}  &  \scalebox{1.1}{6.2}  & \scalebox{1.1}{6.07} & \scalebox{1.1}{8.76}     \\
\scalebox{1.2}{$A_\lambda$} \scalebox{1.2}(GeV)  & \scalebox{1.1}{-$3885.3$}  &  \scalebox{1.1}{$-4995.3$} &  \scalebox{1.1}{$-3192.4$}    & \scalebox{1.1}{$-1414.6$} & \scalebox{1.1}{$-3343.9$}    \\
\scalebox{1.2}{$A_\kappa$} \scalebox{1.1}{(GeV)}  & \scalebox{1.1}{31.1}  &  \scalebox{1.1}{61.1} &  \scalebox{1.1}{40.5} & \scalebox{1.1}{296.4} & \scalebox{1.1}{$-25.2$} \\
\scalebox{1.2}{$\mueff$} \scalebox{1.1}{(GeV)}   & \scalebox{1.1}{$-418.5$}    &  \scalebox{1.1}{$-700.5$}   &   \scalebox{1.1}{$-526.9$}  & \scalebox{1.1}{$-307.2$} & \scalebox{1.1}{$-198.7$}   \\
\scalebox{1.2}{$M_1$} \scalebox{1.1}{(GeV)}    &  \scalebox{1.1}{66.4}    &  \scalebox{1.1}{66.2}  &  \scalebox{1.1}{$-91.67$}  & \scalebox{1.1}{509.2} & \scalebox{1.1}{$-350.1$}    \\
\scalebox{1.2}{$M_2$} \scalebox{1.1}{(GeV)}   &  \scalebox{1.1}{2500}     &  \scalebox{1.1}{2500}  &  \scalebox{1.1}{2500}  & \scalebox{1.1}{2500} & \scalebox{1.1}{2000}    \\
\scalebox{1.2}{$M_{\widetilde{Q}_3}$} \scalebox{1.1}{(GeV)}    &  \scalebox{1.1}{4456.1}    & \scalebox{1.1}{2556.1}  &  \scalebox{1.1}{2560.8}   & \scalebox{1.1}{2217.0} & \scalebox{1.1}{2302.8} \\ 
\scalebox{1.2}{$M_{\widetilde{t}_R}$} \scalebox{1.1}{(GeV)}   &  \scalebox{1.1}{4916.5}   &  \scalebox{1.1}{4916.5}  &   \scalebox{1.1}{4456.0}   & \scalebox{1.1}{4182.1} & \scalebox{1.1}{2569.7}  \\
\scalebox{1.2}{$A_{t}$} \scalebox{1.2} {(GeV)}     &  \scalebox{1.1}{4945.9}   &  \scalebox{1.1}{4945.9}  & \scalebox{1.1}{4526.3} &  \scalebox{1.1}{5126.4}  & \scalebox{1.1}{4868.9} \\ [0.05cm]
\hline
\Tstrut
\scalebox{1.1}{$\mntrlone$} \scalebox{1.2} {(GeV)}   &  \scalebox{1.1}{$-55.5$}   &  \scalebox{1.1}{$-56.8$}  &  \scalebox{1.1}{$-61.7$}  & \scalebox{1.1}{$-295.9$} & \scalebox{1.1}{187.9}  \\
\scalebox{1.1}{$\mntrltwo$}\scalebox{1.2} {(GeV)}    &  \scalebox{1.1}{66.0} &  \scalebox{1.1}{65.8}  & \scalebox{1.1}{$-92.2$} & \scalebox{1.1}{312.2} & \scalebox{1.1}{$-199.5$}  \\
\scalebox{1.1}{$\mntrlthree$} \scalebox{1.1}{(GeV)}     &  \scalebox{1.1}{433.08}  &  \scalebox{1.1}{716.46}  &  \scalebox{1.1}{542.0} & \scalebox{1.1}{$-321.2$} & \scalebox{1.1}{205.7}    \\
\scalebox{1.1}{$\mntrlfour$} \scalebox{1.1}{(GeV)}    &  \scalebox{1.1}{$-435.8$}  &  \scalebox{1.1}{$-719.4$}  &  \scalebox{1.1}{$-544.3$}  & \scalebox{1.1}{509.9} & \scalebox{1.1}{$-356.5$}    \\
\scalebox{1.1}{$m_{\chi_5^0}$} \scalebox{1.1}{(GeV)}    &  \scalebox{1.1}{2553.0}  &  \scalebox{1.1}{2545.4}  &  \scalebox{1.1}{2541.6}  & \scalebox{1.1}{2533.4} & \scalebox{1.1}{2045.2}    \\
\scalebox{1.1}{$\mcharone$} \scalebox{1.1}{(GeV)}  &  \scalebox{1.1}{432.0} &  \scalebox{1.1}{716.5}  &  \scalebox{1.1}{540.6} & \scalebox{1.1}{$-316.7$}  & \scalebox{1.1}{$-205.7$}   \\
\scalebox{1.1}{$m_{h_1}$} \scalebox{1.1}{(GeV)}         &  \scalebox{1.1}{48.8} &  \scalebox{1.1}{50.4}  &  \scalebox{1.1}{70.7} & \scalebox{1.1}{124.1}  & \scalebox{1.1}{125.3}    \\
\scalebox{1.1}{$m_{h_2}$} \scalebox{1.1}{(GeV)}         &  \scalebox{1.1}{124.2}  &  \scalebox{1.1}{124.6}   &  \scalebox{1.1}{123.9}  & \scalebox{1.1}{202.0} & \scalebox{1.1}{175.9}   \\
\scalebox{1.1}{$m_{\as}$} \scalebox{1.1}{(GeV)}        &  \scalebox{1.1}{50.6}   &  \scalebox{1.1}{71.4}   &  \scalebox{1.1}{64.0}   & \scalebox{1.1}{35.9} & \scalebox{1.1}{82.8} \\
\scalebox{1.1}{$m_{H^{\pm}}$} \scalebox{1.1}{(GeV)}       &  \scalebox{1.1}{4074.5}  &  \scalebox{1.1}{5026.3}  & \scalebox{1.1}{3308.4}  & \scalebox{1.1}{1763.5} & \scalebox{1.1}{2419.0}  \\ [0.1cm]
\hline
\Tstrut
\scalebox{1.0}{$N_{11}$}  & \scalebox{1.0}{$0.99$}  & \scalebox{1.0}{$0.99$}   & \scalebox{1.0}{$0.99$}  & \scalebox{1.0}{$0.96$} & \scalebox{1.0}{$0.99$}    \\
\scalebox{1.0}{$N_{21}$}  & \scalebox{1.0}{$-0.99$}  & \scalebox{1.0}{$-0.99$}   & \scalebox{1.0}{$0.99$}  & \scalebox{1.0}{$-0.1$} & \scalebox{1.0}{$-0.2$}    \\
\hline
\scalebox{1.1}{$\Omega h^2$} & \scalebox{1.1}{0.121} &  \scalebox{1.1}{0.118}  &  \scalebox{1.1}{0.124}  & \scalebox{1.1}{0.119} & \scalebox{1.1}{0.122}   \\[0.10cm]
\scalebox{1.1}{$\sigma^{\rm SI}_{\chi^0_1-p(n)}$} \scalebox{1.1}{(cm$^2$)}   &    \scalebox{1.1}{$2.9(2.8)\times 10^{-48}$}  &    \scalebox{1.1}{$1.2(1.3)\times 10^{-48}$}  & \scalebox{1.1}{$5.6(6.0)\times 10^{-48}$}   &  \scalebox{1.1}{$8.5(8.6)\times 10^{-47}$}   &  \scalebox{1.1}{$3.15(3.2)\times 10^{-48}$}    \\[0.30cm]
\scalebox{1.1}{$\sigma^{\rm SD}_{\chi^0_1-p(n)}$} \scalebox{1.1}{(cm$^2$)}    &    \scalebox{1.1}{$9.03(-6.9)\times 10^{-44}$}  &    \scalebox{1.1}{$1.1(0.9)\times 10^{-44}$}    & \scalebox{1.1}{$7.5(5.7)\times 10^{-43}$}  &  \scalebox{1.1}{$3.4(2.6)\times 10^{-42}$} &  \scalebox{1.1}{$8.12(-6.2)\times 10^{-44}$}   \\[0.15cm]
\hline
\scalebox{1.1}{BR($\chi^\pm_1\to\chi_1^0 W^\pm $)}  &  \scalebox{1.1}{0.13}  &  \scalebox{1.1}{0.13}  &  \scalebox{1.1}{0.43}  & \scalebox{1.1}{0.00} & \scalebox{1.1}{0.00} \\[0.1cm]
\scalebox{1.1}{BR($\chi^\pm_1\to\chi_2^0 W^\pm $)}  &  \scalebox{1.1}{0.87}   &  \scalebox{1.1}{0.87} &  \scalebox{1.1}{0.57}   & \scalebox{1.1}{0.00}  & \scalebox{1.1}{0.00} \\[0.05cm]
\scalebox{1.1}{BR($\chi^\pm_1\to\chi_1^0 f \bar{f} $)}  &  \scalebox{1.1}{0.00} &  \scalebox{1.1}{0.00} &  \scalebox{1.1}{0.00} & \scalebox{1.1}{0.99}  & \scalebox{1.1}{0.80} \\[0.1cm]
\scalebox{1.1}{BR($\chi^\pm_1\to\chi_2^0 f \bar{f} $)}  &  \scalebox{1.1}{0.00} &  \scalebox{1.1}{0.00} &  \scalebox{1.1}{0.00}  & \scalebox{1.1}{0.01} & \scalebox{1.1}{0.20}  \\[0.08cm]
\hline
\scalebox{1.1}{BR($\chi^0_2\to\chi_1^0 f \bar{f} $)}  &  \scalebox{1.1}{0.005} &  \scalebox{1.1}{0.12}  & \scalebox{1.1}{0.28} &  \scalebox{1.1}{0.37}  & \scalebox{1.1}{0.23} \\[0.05cm]
\scalebox{1.1}{BR($\chi^0_2\to\chi_1^0 \gamma $)}  &  \scalebox{1.1}{0.995}  &  \scalebox{1.1}{0.88} & \scalebox{1.1}{0.72} &  \scalebox{1.1}{0.63} & \scalebox{1.1}{0.73}   \\[0.05cm]
\hline
\scalebox{1.1}{BR($\chi^0_3\to\chi_1^0 Z$)}   &  \scalebox{1.1}{0.06}  &  \scalebox{1.1}{0.05} &  \scalebox{1.1}{0.18}  & \scalebox{1.1}{0.00} & \scalebox{1.1}{0.00}  \\[0.07cm]
\scalebox{1.1}{BR($\chi^0_3\to\chi_1^0 \, \hsm$)}  &  \scalebox{1.1}{0.06}  &  \scalebox{1.1}{0.07} &  \scalebox{1.1}{0.22} & \scalebox{1.1}{0.00} & \scalebox{1.1}{0.00} \\[0.05cm]
\scalebox{1.1}{BR($\chi^0_3\to\chi_2^0 Z$)}   &  \scalebox{1.1}{0.42}  &  \scalebox{1.1}{0.48}  &  \scalebox{1.1}{0.30} & \scalebox{1.1}{0.00} & \scalebox{1.1}{0.00}  \\[0.05cm]
\scalebox{1.1}{BR($\chi^0_3\to\chi_2^0 \, \hsm$)}  &  \scalebox{1.1}{0.45} &  \scalebox{1.1}{0.40}   &  \scalebox{1.1}{0.28}  & \scalebox{1.1}{0.00} & \scalebox{1.1}{0.00}  \\[0.05cm]
%
%\scalebox{1.1}{BR($\chi^0_3\to\chi_1^0 \, f \bar{f}$)}  &  \scalebox{1.1}{0.00}  &  \scalebox{1.1}{0.00}  &  \scalebox{1.1}{0.00}  & \scalebox{1.1}{0.02} & \scalebox{1.1}{0.002} \\[0.05cm]
%
\scalebox{1.1}{BR($\chi^0_3\to\chi_2^0 \, f \bar{f}$)}  &  \scalebox{1.1}{0.00}  &  \scalebox{1.1}{0.00} &  \scalebox{1.1}{0.00} & \scalebox{1.1}{0.12}  & \scalebox{1.1}{0.078} \\[0.05cm]
%
%\scalebox{1.1}{BR($\chi^0_3\to\chi_1^0 \gamma$)}   &  \scalebox{1.1}{0.00}  &  \scalebox{1.1}{0.00} & \scalebox{1.1}{0.00}  &  \scalebox{1.1}{0.001}  & \scalebox{1.1}{0.0001} \\[0.05cm]
%
\scalebox{1.1}{BR($\chi^0_3\to\chi_2^0 \gamma$)}   &  \scalebox{1.1}{0.00}  &  \scalebox{1.1}{0.00}  & \scalebox{1.1}{0.00}  &  \scalebox{1.1}{0.86} &  \scalebox{1.1}{0.92} \\[0.07cm]
\hline
\Tstrut
\scalebox{1.1}{BR($\chi^0_4\to\chi_1^0 Z$)}    &  \scalebox{1.1}{0.08} &  \scalebox{1.1}{0.08}  &  \scalebox{1.1}{0.26} & \scalebox{1.1}{0.02} & \scalebox{1.1}{0.006}  \\[0.05cm]
\scalebox{1.1}{BR($\chi^0_4\to\chi_1^0 \, \hsm$)} &  \scalebox{1.1}{0.05} &  \scalebox{1.1}{0.05}  &  \scalebox{1.1}{0.16} & \scalebox{1.1}{0.002} & \scalebox{1.1}{0.0001}  \\[0.05cm]
\scalebox{1.1}{BR($\chi^0_4\to\chi_2^0 Z$)}   &  \scalebox{1.1}{0.49}  &  \scalebox{1.1}{0.38}   &  \scalebox{1.1}{0.29}  & \scalebox{1.1}{0.04} & \scalebox{1.1}{0.02} \\[0.05cm]
\scalebox{1.1}{BR($\chi^0_4\to\chi_2^0 \, \hsm$)}  &  \scalebox{1.1}{0.37}  &  \scalebox{1.1}{0.49}  &  \scalebox{1.1}{0.28}  &  \scalebox{1.1}{0.23} & \scalebox{1.1}{0.2} \\[0.05cm]
\scalebox{1.1}{BR($\chi^0_4\to\chi_3^0 \, Z$)}  &  \scalebox{1.1}{0.00}  &  \scalebox{1.1}{0.00}  &  \scalebox{1.1}{0.00} &  \scalebox{1.1}{0.18}  & \scalebox{1.1}{0.21} \\[0.05cm]
\scalebox{1.1}{BR($\chi^0_4\to\chi_3^0 \, \hsm$)}  &  \scalebox{1.1}{0.00}   &  \scalebox{1.1}{0.00}  &  \scalebox{1.1}{0.00} &  \scalebox{1.1}{0.13} & \scalebox{1.1}{0.003}  \\[0.05cm]
\scalebox{1.1}{BR($\chi^0_4 \to \charonepm \, W^{\mp}$)}   &  \scalebox{1.1}{0.00}  &  \scalebox{1.1}{0.00}  &  \scalebox{1.1}{0.00}  &  \scalebox{1.1}{0.52} &  \scalebox{1.1}{0.54} \\[0.05cm]
\hline
\Tstrut
\scalebox{1.1}{$\sigma_{pp \rightarrow \ntrltwothreefour \charonepm}$~(pb)}  &  \scalebox{1.1}{0.0418}   &  \scalebox{1.1}{0.00425}  &  \scalebox{1.1}{0.01577}  &  \scalebox{1.1}{0.140} &  \scalebox{1.1}{0.743} \\[0.07cm]
\hline 
\scalebox{1.1}{\texttt{CheckMATE} result}  &  \scalebox{1.1}{Excluded}  &  \scalebox{1.1}{Allowed}    & \scalebox{1.1}{Allowed} &  \scalebox{1.1}{Allowed} & \scalebox{1.1}{Allowed} \\
\scalebox{1.1}{$r$-value}           &  \scalebox{1.1}{2.87}     &  \scalebox{1.1}{0.68}     & \scalebox{1.1}{0.61} & \scalebox{1.1}{0.07} & \scalebox{1.1}{0.12} \\
\scalebox{1.1}{Analysis ID}        &  \scalebox{1.0}{atlas$\_$2004$\_$10894}\cite{ATLAS:2020qlk} &  \scalebox{1.0}{atlas$\_$2004$\_$10894}\cite{ATLAS:2020qlk} & \scalebox{1.0}{atlas$\_$2004$\_$10894}\cite{ATLAS:2020qlk}  & \scalebox{1.0}{atlas$\_$conf$\_$2017$\_$060}\cite{ATLAS:2017dnw} & \scalebox{1.0}{atlas$\_$conf$\_$2020$\_$048}\cite{ATLAS:2020wzf}\\
\scalebox{1.1}{Signal region ID} &  \scalebox{1.1}{Cat12}  &  \scalebox{1.1}{Cat12}  & \scalebox{1.1}{Cat12} &  \scalebox{1.1}{EM7} & \scalebox{1.1}{EM09} \\[0.05cm]\hline
\end{tabular}
}}
%\vskip 5pt
\caption{Benchmark scenarios allowed (BP1 to BP4) by all relevant theoretical and 
experimental constraints and disallowed (BP-D1) by the LHC searches for the electroweakinos. Various input parameters and the resulting masses, mixings, and branching fractions of the relevant states, along with 
the values of various DM observables, are provided. The most sensitive LHC analyses with their signal regions that are studied using \checkmate~are presented at the end. We do not present the \smodels~results for these benchmark points as they exhibit lower sensitivity.}
\label{tab:benchmarks}
}
%\end{center}
\end{table}

In contrast to the co-annihilation scenario in BP1, BP2 features a scenario where singlino-dominated DM annihilation through the resonant s-channel $\hsm$ exchange, leading to a rapid reduction in the DM relic density.
This resonance phenomenon helps comply with the Planck-observed upper bound on the DM relic abundance.
While the nearby bino state has a limited impact on estimating relic abundance, it significantly influences the DMDD-SI cross-section and collider searches for DM. 
The presence of $M_1 \sim -90$~GeV tempers the singlino-higgsino-bino system, reducing the DMDD-SI cross-section of the singlino-dominated DM to approximately $10^{-48} \text{cm}^2$. 
Additionally, beyond its impact on the DMDD-SI cross-section, such a bino state leaves its imprint on collider searches for DM. 
BR$[\ntrltwo \rightarrow \ntrlone \gamma]$ is around 72\%.  The scenario parallels BP1, where higgsino-like $\ntrlthreefour$ and $\charonepm$ undergo significant cascade decay through $\ntrltwo$. However, given that the mass difference $\mntrltwo - \mntrlone \sim 30$~GeV is relatively larger than in the BP1 scenario, the radiated photon here is expected to be relatively hard compared to BP1. 
On the other hand, note that $\mueff$ in BP2 is much smaller than that of BP1, resulting in the production cross-section of the process $pp \rightarrow \ntrlthreefour \charonepm$ being almost four times larger than BP1. Nevertheless, a \checkmate~analysis allows this benchmark point with a smaller $r$ ($= 0.61$) value compared to BP1. This can be understood from the fact that, in BP2, $\lambda$ is almost twice of BP1, which reduces the branching fractions of higgsino-like states to decay to the bino-like NLSP, resulting in a reduction in the yields of final states with photons. This drops the sensitivity of the ATLAS analysis~\cite{ATLAS:2020qlk} for this benchmark point.

We present another benchmark point, BP3, where the observed relic abundance of DM is realized via singlino higgsino co-annihilation.
The relative contribution to the relic density of the annihilation processes of  $\{\charonepm \ntrltwothree \rightarrow \text{SM}  \, \,  \text{SM}\}$ and $\{\charonepm \charonemp \rightarrow \text{SM} \, \, \text{SM}\}$ are much larger than the co-annihilation process involving LSP, such as, $\{\charonepm \ntrlone \rightarrow \text{SM}  \, \,  \text{SM}\}$.
Thus, in this scenario, “assisted co-annihilation”~\cite{Ellwanger:2009dp} plays a very important role in reconciling the relic density of the singlino-dominated DM with the Planck measurement.
The spectrum of the singlino higgsino sector corresponds to a compressed scenario. Note that, here singlino-dominated DM and the higgsino-like chargino/neutralinos are around 300 GeV, which seems to be much above the latest constraints from LHC analyses~\cite{ATLAS:2021moa}. 
It is important to note that, in this scenario, such experimental bounds would be relaxed significantly due to the smaller production cross-section (smaller
by 50\%–60\%) of higgsino-like $\ntrltwothree \charonepm$ when compared to that of wino-like $\ntrltwo \charonepm$ assumed in the LHC analyses.
Furthermore, these experimental analyses consider the wino-like states decay via off-shell $W, Z, \hsm$ states whereas in this benchmark point $\ntrltwothree$ exhibit significant amount of radiative decays which significantly suppress the reach of the multi-lepton search analyses at the LHC.

Here, BR$[\ntrltwo \rightarrow \ntrlone \gamma]$  and BR$[\ntrlthree \rightarrow \ntrltwo\gamma]$ are around 63\% and 86\%, respectively. One photon would appear from the decay of $\ntrltwo$, and two photons would appear from the cascades of $\ntrlthree$.
Note that, $\mntrltwo - \mntrlone \sim 16$ GeV, $\mntrlthree - \mntrlone \sim 25$ GeV and $\mntrlthree - \mntrltwo \sim 9$ GeV. Thus, in this scenario, the photons coming from the higgsino-like neutralinos and other visible decay products of the higgsino-like chargino are expected to be relatively soft to be detected.
The current investigations at the LHC do not examine this specific area of parameter space, mostly because of the large radiative branching ratios of higgsino-like heavier neutralinos. This impedes the decay into charged-lepton final states, which are predominantly considered in the current LHC searches.
Therefore, it indicates the possibility of utilizing a new detection channel at the LHC more effectively to explore this region of the NMSSM parameter space.
To augment the distinctive features of the radiative decay channel involving a soft photon and $\etmiss$ at the LHC, one may explore events where higgsino pairs ($\ntrltwothree + \charonepm$) are generated alongside an initial state radiation (ISR) jet (j) which is relatively hard~\cite{Baum:2023inl}.
Despite the limited impact of the ISR jet on the distribution of $P_T$  of the photon, it actively recoils against the ($\ntrltwothree + \charonepm$) system, which leads to a sizable increase $\etmiss$ of such events. A comprehensive collider simulation and discussion on this is presented in the following subsection.

At this benchmark point, the bino-like fourth neutralino state indirectly impacts both collider and DMDD phenomena. A mass parameter $M_1 \sim 509$~GeV introduces a splitting of approximately 10 GeV between the higgsino-like states $\ntrltwo$ and $\ntrlthree$. This splitting leads to a notable decay branching fraction of $\ntrlthree$ into $\ntrltwo$ and $\gamma$. It is noteworthy that the sign of $M_1$ plays a crucial role. Such a splitting between the higgsino-like states $\ntrltwo$ and $\ntrlthree$ is unattainable for $M_1 \sim  -509$~GeV, leading to a shift in the decay patterns of both $\ntrltwo$ and $\ntrlthree$ with a significant reduction in radiative decay branching fractions.
The DMDD-SI cross-section remains below the latest experimental bounds due to $M_1 > 0$. If we consider $M_1 < 0$ with the same absolute value, the DMDD-SI cross-section would be an order of magnitude larger, leading to exclusion from the latest DMDD-SI constraints. 
Hence, as previously discussed, the bino state, serving as the fourth neutralino, continues to impact the exploration of singlino-dominated DM in both collider and DMDD experiments.

Benchmark point BP4 shares similarities with BP3 in its singlino-higgsino co-annihilation scenario but stands out with a relatively small $\mu_{\text{eff}}$ of about $200$~GeV, leading to a significantly augmented production cross-section of higgsino-like electroweakinos at the LHC. Here also, both the higgsino-like neutralinos ($\ntrltwothree$)  exhibit large radiative decay modes in the singlino-dominated LSP and a photon rather than decay
into leptons/hadrons. Thus, BR$[\ntrltwo \rightarrow \ntrlone \gamma]$  and BR$[\ntrlthree \rightarrow \ntrltwo\gamma]$ are around 73\% and 92\%, respectively.
 However, the mass gaps between the singlino-like LSP and the higgsino-like neutralinos are relatively small compared to BP3. Consequently, it is expected that the photons arising from the decay of higgsino-like states would be relatively soft compared to those in BP3. In this benchmark point, since the $\kappa < 0$, $M_1$ and $\mueff$ have same relative sign, i.e., both are $-$ve, which reduces the DMDD-SI cross-section below the latest experimental constraints.
Concerning collider studies, an analysis using \checkmate, which includes all available analyses in its repository, reveals an 'r' value significantly below 1 for points BP3 and BP4. This indicates a notable insensitivity to LHC searches, implying that these points are allowed.

It is interesting to note that both ATLAS~\cite{ATLAS:2021moa, ATLAS:2019lng} and CMS~\cite{CMS:2021cox, CMS:2021edw} collaborations observe small excess in searches for the chargino-neutralinos in the compressed regions in the soft di-lepton channels.
This excess can be interpreted as the production of higgsino-like states where the mass splitting between the chargino and the LSP is around $5–20$ GeV~\footnote{Note that, all the reported excesses have the local significance $\lesssim 2.5\sigma$~\cite{ATLAS:2019lng, CMS:2021edw, ATLAS:2021moa}. Recently, in reference~\cite{Agin:2023yoq}, it has been claimed that the excess can also be realised from the monojet searches~\cite{ATLAS:2021kxv,CMS:2021far} in a similar region of parameter space in the context of light-compressed higgsinos after recasting those analyses in Madanalysis5~\cite{Araz:2020lnp}.}. Note that the excess in the soft lepton channels can be explained within the context of singlino-higgsino co-annihilation scenarios discussed in this work. As we pointed out, such a co-annihilation scenario can also indicate another possible detection channel involving photons.
 Therefore, if the LHC conducts a dedicated analysis using existing Run-2 data to probe this scenario with photons, it would be intriguing to observe whether the same excess is present. Nevertheless, these compressed regions will undergo further investigation in the upcoming runs of the LHC, and it will be interesting to see whether the excess persists.

\subsection{Possible new search channels at the LHC }
In this section, we take a closer look at the differential distributions of various kinematic variables and their correlations in order to differentiate signal events 
from backgrounds.
To analyze the kinematics of our signal events of the benchmark points, 
we simulate $pp \to \ntrltwothreefour \charonepm$
events for all benchmark points at the
lowest order (LO) in the perturbation theory using
{\tt MadGraph5-aMC$@$NLO-v2.7.3}~\cite{Alwall:2014hca} for the 14 TeV 
HL-LHC run. 
The default parton distribution function ({\tt `nn23lo1'}~\cite{Ball:2010de}) is utilized with a dynamic factorization scale selection defined as $Q^2= \frac{1}{2} \sum_i M_{i}^{T^2} =\frac{1}{2}\sum_i(m_i^2+p_{Ti}^2)$, where $M^T_i$ represents the transverse mass of the '$i$'-th final state particle.
We focus on studying the distribution of kinematical variables such as $\etmiss$, $P_T$ of the leading jet $(P_T^{\text{jet}})$  and photon $(P_T^{\gamma})$.
Particularly for BP3 and BP4, we also simulate $pp \to \ntrltwothreefour \charonepm j$ process to study the effect of the ISR on distributions of these kinematic variables.
To generate signal events, the interface between {\tt MadGraph5} and the NMSSM is facilitated by incorporating the {\tt UFO} model file~\cite{Degrande:2011ua} from the {\tt FeynRules} package~\cite{Alloul:2013bka}.

Calculations for the decay kinematics of unstable excitations, as well as the subsequent processes of showering and hadronization, are executed employing {\tt PYTHIA8-v8.3}~\cite{Sjostrand:2006za, Sjostrand:2007gs}. The default \pythia8 cards are employed for these simulations.
As mentioned earlier, various masses and decay branching fractions of NMSSM excitations of those benchmark points are generated via {\nmssmtools} and \pythia8 reads those from the {\nmssmtools-generated} SLHA \cite{Skands:2003cj} files.
Finally, the effects of the detector response are incorporated using {\tt DELPHES-v3.4.2}~\cite{deFavereau:2013fsa}.
Utilizing Prospino 2.1~\cite{Beenakker:1996ed}, we compute the NLO corrections for electroweakino production, projecting a 20-30\% increase in the LHC cross-sections within the region of parameter space of present interest. To accommodate this, we apply a common K-factor of 1.25~\cite{Fiaschi:2018hgm}.
%

%%%%%%%%%%%%%%%%%%
%
\begin{figure}[htbp]
\begin{center}
\includegraphics[height=5.3cm,width=0.32\linewidth]{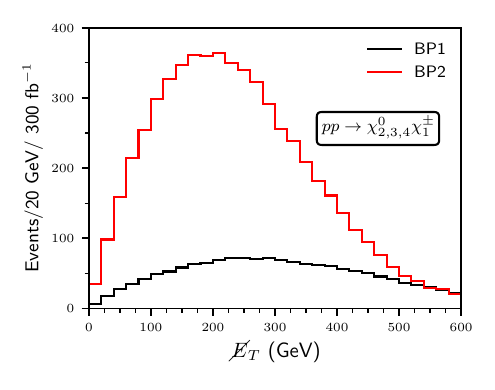}
\hskip 5pt
\includegraphics[height=5.3cm,width=0.32\linewidth]{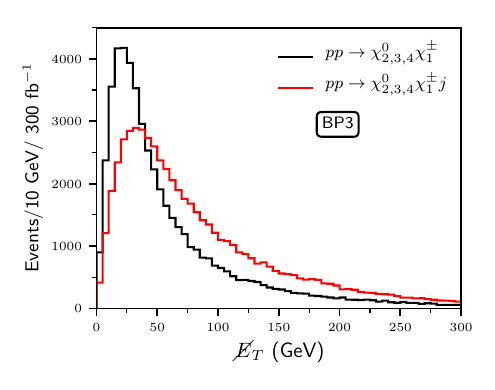}
\hskip 1pt
\includegraphics[height=5.3cm,width=0.32\linewidth]{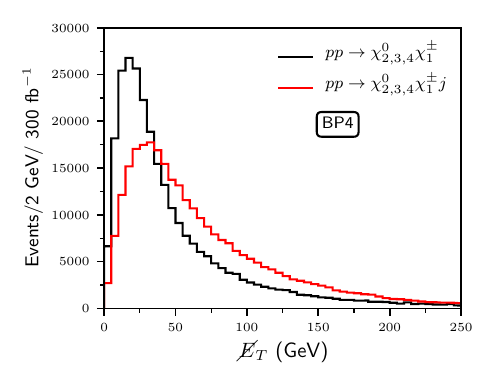}
\vskip 1pt
\includegraphics[height=5.3cm,width=0.32\linewidth]{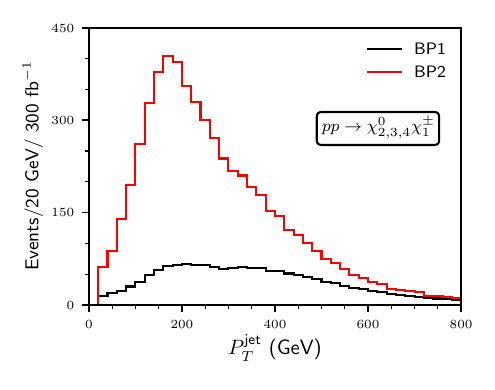}
\hskip 1pt
\includegraphics[height=5.3cm,width=0.32\linewidth]{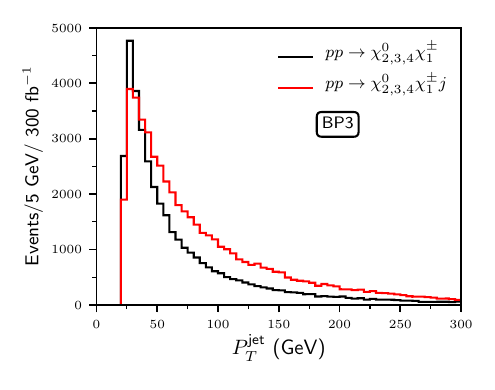}
\hskip 1pt
\includegraphics[height=5.3cm,width=0.32\linewidth]{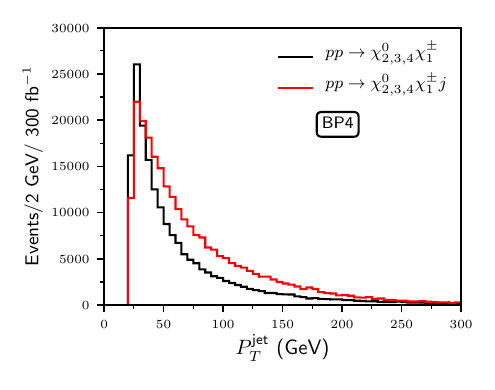}
\vskip 1pt
\includegraphics[height=5.3cm,width=0.32\linewidth]{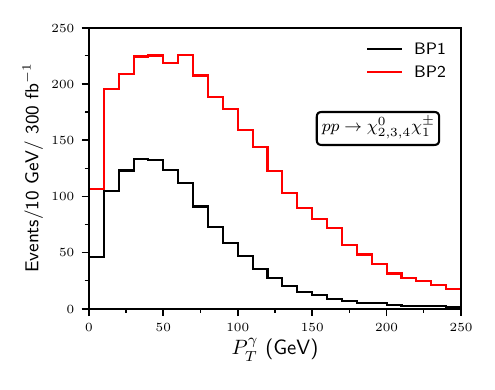}
\hskip 1pt
\includegraphics[height=5.3cm,width=0.32\linewidth]{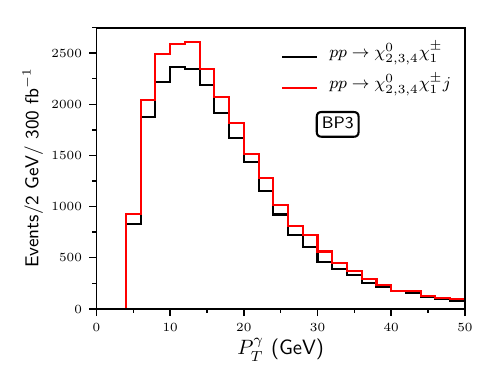}
\hskip 1pt
\includegraphics[height=5.3cm,width=0.32\linewidth]{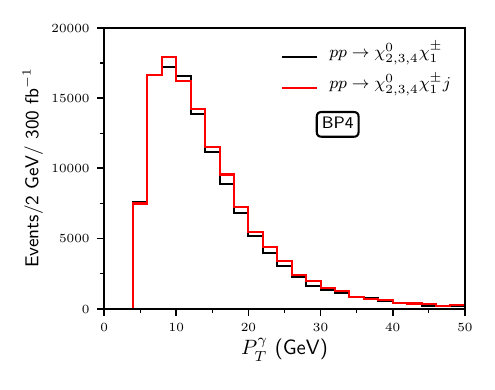}
\caption{Differential event distributions of $\etmiss$, $P_T^{\text{jet}}$, and $P_T^{\gamma}$ from top to bottom in the left panel for BP1 (in black) and BP2 (in red). On the middle and the right panel, differential event distributions for similar kinematic variables are presented for BP3 and BP4, respectively, where black (red) indicates distributions for the process $pp \to \ntrltwothreefour \charonepm$ ($pp \to \ntrltwothreefour \charonepm j$).}
\label{fig:bp123distributions}
\end{center}
\vspace{-0.5cm}
\end{figure}
%%%%%%%%%%%%
%

As discussed in the preceding section, for BP1 and BP2, the production cross-section of the process $pp \rightarrow \ntrltwo \charonepm$ is significantly smaller, given the fact that $\ntrltwo$ is a bino-dominated neutralino, in contrast to the production process $pp \rightarrow \ntrlthreefour \charonepm$, where $\ntrlthreefour$ are higgsino-dominated neutralinos.
In the left panel of figure~\ref{fig:bp123distributions}, the top-to-bottom sequence displays the distributions of $\etmiss$, $P_T^{\text{jet}}$ and $P_T^{\gamma}$, respectively. The distributions in black (red) are for BP1 (BP2).
Relatively lighter higgsino-like states (smaller $\mueff$) in BP2 result in a larger production cross-section compared to BP1, leading to significantly larger amplitudes of events in its differential distributions of those kinematic
variables.
Since both the benchmark points exhibit a similar type of hierarchy among the states, the distributions of $\etmiss$ and $P_T^{\text{jet}}$ peak around similar ranges. On the other hand, as expected, due to a relatively larger mass gap between $\ntrltwo$ and $\ntrlone$, the distribution of $P_T^{\gamma}$ peaks at a relatively larger value for BP2 compared with BP1.
Therefore, in order to suppress the SM background, one can demand a selection cut on a relatively larger value of $\etmiss$ ($\gtrsim 150$~GeV) along with a similarly relatively larger $P_T$ of the leading jets/leptons coming from on-shell $W, Z, \hsm$-bosons resulting from the decay of
heavier higgsino-like states ($\ntrlthreefour, \charonepm$). Additionally, implementing event selection cuts on the $P_T$ of the relatively soft leading photon ($\gtrsim 10$ GeV) can further enhance the suppression of the SM background. 
Therefore, as we mentioned in the previous section, integrating selection criteria that include hard jets/leptons, large missing energy, and a relatively soft photon in the analysis would facilitate the search of the singlino-bino co-annihilation scenario with relatively large higgsino-like states.

In the middle and the right panel of Figure~\ref{fig:bp123distributions}, the sequence from top to bottom illustrates distributions of the same kinematic variables for BP3 and BP4, respectively. The black and red curves in these plots represent the processes $pp \to \ntrltwothreefour \charonepm$ and $pp \to \ntrltwothreefour \charonepm j$, respectively. 
Relatively smaller $\mu_{\text{eff}}$ in BP4 results in a greater production cross-section of the higgsino-like states compared to BP3. This, in turn, leads to significantly larger amplitudes of events in the differential distributions of those kinematic variables.
In the overall production cross-section for both BP3 and BP4, the process involving $\ntrlfour$, being bino-like, is subdominant.
As discussed in the previous section, unlike BP1 and BP2, BP3 and BP4 feature singlino-higgsino compressed scenarios where higgsino-like neutralinos ($\ntrltwothree$) prominently exhibit radiative decay modes, favoring emissions involving the singlino-dominated LSP and a photon rather than decay into leptons/hadrons. Additionally, due to the limited available phase space, the visible decay productions from the higgsino-like states are expected to be relatively soft. In order to suppress the SM background, events featuring the production of higgsino-like states ($\ntrltwothree \charonepm$) in conjunction with a hard ISR jet can be considered.
In the absence of an ISR jet, $\ntrltwothree$ and $\charonepm$ would primarily be produced at the LHC with equal and opposite $P_T$. However, in the presence of the ISR jet, the ($\ntrltwothree \charonepm$) system recoils against the ISR jet in the transverse plane.
Due to the small mass difference between the LSP and the higgsino-like states, a significant portion of the $P_T$ of the higgsino-like $\ntrltwothree$ and $\charonepm$ is transferred to the LSP, contributing to event $\etmiss$ that approximately balances with $P_T$ of the ISR jet.
The plots reveal that the peak of $\etmiss$ distribution occurs at a relatively higher value for the process involving the ISR jet. Additionally, a broad high $\etmiss$ tail is observed for events containing one ISR jet. This characteristic allows for more aggressive selection cuts on $\etmiss$ in the analysis, effectively rejecting a significant amount of the SM backgrounds at a moderate cost in losing signal events.

It is noteworthy that a similar broader high $P_T$ tail of the leading jet is also observed in events containing one ISR jet. Given the correlation between $P_T^{\text{jet}}$ and $\etmiss$ in events with one ISR jet, imposing a stringent cut on $\etmiss$ ($\gtrsim$ 100 GeV) ensures that most signal events have substantially larger $P_T^{\text{jet}}$ ($\gtrsim$ 100 GeV). Consequently, incorporating hard cuts on both $\etmiss$ and $P_T^{\text{jet}}$ enables a significant reduction in the SM background while minimally sacrificing signal events.
On the other hand, distributions of the $P_T$ of the leading photon for events with and without an ISR jet appear nearly identical. Due to comparatively larger available phase space for the decay of $\ntrltwothree$ to $\ntrlone$, the distribution of $P_T$ of the leading photon exhibits a relatively broader range, with the peak occurring at a higher value in BP3 as compared to BP4.

%%%%%%%%%%%%%%%%%%
%
\begin{figure}[t!]
\begin{center}
\includegraphics[height=5.9cm,width=0.49\linewidth]{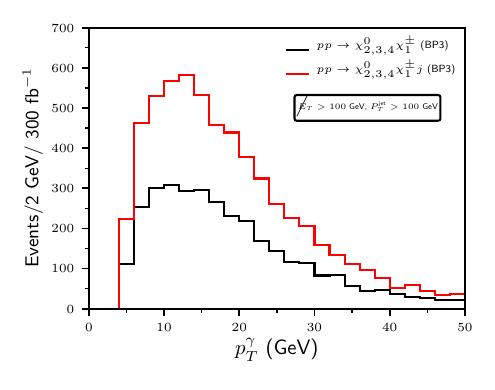}
\hskip 1pt
\includegraphics[height=5.9cm,width=0.49\linewidth]{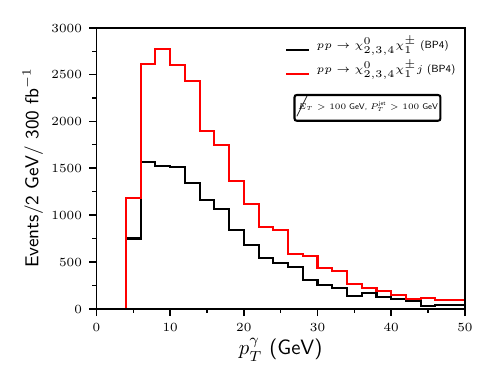}
\caption{The distribution of $P_T$ of the leading photon for BP3 (left) and BP4 (right) considering $\etmiss$ and $P_T$ of the leading jet more than 100 GeV. Black and red denote event distributions for $pp \to \ntrltwothreefour \charonepm$ and $pp \to \ntrltwothreefour \charonepm j$ processes, respectively.}
\label{fig:bp3ptgammaaftercuts}
\end{center}
\vspace{-0.5cm}
\end{figure}
%%%%%%%%%%%%
%

In figure~\ref{fig:bp3ptgammaaftercuts}, we present the event distribution of $P_T^{\gamma}$ for BP3 (left) and BP4 (right) considering the two above mentioned cuts, i.e., $\etmiss > 100$~GeV and $P_T^{\text{jet}} > 100$~GeV. 
The presence of a single ISR jet in the events under those specified cuts leads to a notable increase in the number of events at the peak of the distribution and a broadening of the high $P_T^{\gamma}$ tail. This phenomenon arises due to a substantial drop in the cross-section of the process in the absence of any ISR jet under such cuts.
Moreover, the distribution exhibits a peak at a slightly higher $P_T^{\gamma}$ when the ISR jet is considered, suggesting an overall transverse boost for the photon in the context of $pp \to \ntrltwothreefour \charonepm j$, particularly for a relatively hard ISR jet.
This can be understood from the fact that if the decaying photon from $\ntrltwothree$ originated in the same direction in which $\ntrltwothree$ are produced and boosted due to large $P_T$ of the ISR jet in the event.
These observations imply a significant influence of the ISR jet on both the yield of events and the kinematics of the final-state photon.
The enhanced peak and the extended high tail of the distribution of $P_T^{\gamma}$ underscore the importance of accounting for a hard ISR jet in the process.
Consequently, triggering on a photon with $P_T > 10$ GeV in such a scenario would capture a significantly larger number of signal events in $pp \to \ntrltwothreefour \charonepm j$ compared to $pp \to \ntrltwothreefour \charonepm$ under the aforementioned cuts on $\etmiss$ and $P_T^{\text{jet}}$.

In this section, we present differential distributions of some kinematic variables and propose some signal regions in terms of these variables for BP1, BP2, BP3 and BP4. 
However, more dedicated analyses are essential for the proposed signals and the SM background to explore novel signatures at the LHC, thereby discerning the most promising signal topologies and optimal event selection strategies.
Conducting multivariate analyses with an expanded set of kinematic variables for all visible objects resulting from the decays of heavier neutralinos (including leading jets, leptons, photons, and missing energy) can become essential in some scenarios. This strategy can be instrumental in significantly reducing the SM background and improving signal efficiency. We preserve this for our future work.
Note that a recent work~\cite{Datta:2022bvg} demonstrates that relatively not-so-heavy top squarks in the NMSSM can probe a unique parameter space of the higgsino-singlino-bino system involving a relatively light singlet-like scalar. 
A relatively larger production cross-section of top squark pairs at the LHC is instrumental in enabling the exploration of this specific parameter space, which, remarkably, has not yet been investigated by experimental collaborations.
In the context of our specific region of interest, characterized by substantial radiative decay modes of heavier neutralinos that involve photon emission, the production of moderately heavy top squarks at the LHC could also yield intriguing collider signals, in a way similar to an analysis in reference~\cite{Baer:2005jq}. 
Higgsino-like or bino-like neutralinos can arise from the decay cascades of top squarks, and these neutralinos can emit a photon. In this context, the photon could exhibit significant boosting compared to the scenario we are discussing, especially if a substantial mass gap exists between the top squark and the decaying neutralino state, resulting in a boosted neutralino. 
A dedicated analysis can be undertaken to investigate such compelling collider signatures arising from the region of interest delineated in this study.
\section{Conclusion}
\label{conclusion}
Understanding the nature of the DM stands out as a prominent challenge in theoretical particle physics and cosmology, necessitating the exploration of novel physics that extends beyond the confines of the SM.
In this work, we investigate the electroweakino sector of the NMSSM, which can provide a cold DM candidate for the Universe. We focus on the singlino-dominated LSP DM, tempered by the nearby higgsino-like and bino-like neutralinos. An involved set of blind spot conditions of DMDD-SI and DMDD-SD scattering cross-sections for the singlino-dominated DM are derived considering $(4\times4)$ bino-higgsino-singlino neutralino sector.
At the minimal mixing between the SM-like Higgs boson and the other Higgs bosons and considering only the singlino-higgsino neutralino sector, a spin-independent blind spot condition arises exclusively when the ratio $\mntrlone/\mueff > 0$, i.e., $\kappa > 0$.
It has been shown in this work that due to the tempering of the bino-like neutralino, a novel blind spot condition can appear for the singlino-dominated DM when $\mntrlone$ and $\mueff$ carry an opposite relative sign, i.e., $\kappa < 0$ region of parameter space. 
Such a scenario arises due to a cancellation between the $g_1$ proportional gaugino-higgsino-Higgs boson interaction term and the $\lambda$ proportional singlino-higgsino-Higgs boson interaction term. 
It demands the same  (opposite) relative sign between $\mueff$ and $M_1$ for $\kappa <0$ $(> 0)$.
This opens up a new region of parameter space that exhibits a smaller DMDD-SI cross-section of the singlino-dominated DM.
On the other hand, the spin-dependent rate (proportional with $g_{\ntrlone \ntrlone Z}$ $\sim N_{13}^2-N_{14}^2$, i.e.,  the difference between the higgsino components of the LSP) is suppressed in the MSSM when $\tanb$ is small (approaching zero in the limit of $\tanb \rightarrow 1$ or very large values of $\mueff$), could now have new compensating terms in its expression in the
presence of a light bino-like state. These can then play their roles in suppressing the
relevant coupling even for larger $\tanb$ values.

In addition to highlighting the role of light bino-like and higgsino-like states in tempering the singlino-dominated LSP on the DMDD rates, we explore scenarios where $M_1$ or $\mueff$ is close to $\msinglino$ and as a result the bino or the higgsino states can act as co-annihilation partners for singlino-dominated DM, aiming to ensure compliance with the observed DM relic abundance of the Universe.
In the case of singlino-bino co-annihilation, NLSP is bino-like, and it can exhibit a significant radiative decay into a soft photon and the singlino-dominated DM. In this scenario, cascades of the produced higgsino-like $\ntrlthreefour$ and $\charonepm$ via $\ntrltwo$ could lead to: $pp \rightarrow \ntrlthreefour (\widetilde{H}) \charonepm (\widetilde{H}) 
\rightarrow $ $\hsm/Z + W^{\pm} + \ntrltwo (\tilde{B}) \, \big[\ntrltwo \rightarrow  \gamma \, \ntrlone (\tilde{S})\big]$
%$\rightarrow $ $\hsm/Z + W^{\pm} + \etmiss$ + $\geq  1\gamma$
$\Rightarrow 3\ell + \geq 1\gamma + \etmiss$ or $1\ell + 2b + \geq 1\gamma + \etmiss$ final states at the LHC.
If the photon from $\ntrltwo$ remains soft and undetected, the cascades of $\charonepm$ and $\ntrlthreefour$ via bino-like $\ntrltwo$ lead to similar final states as when these higgsino-like states directly decay to the singlino-dominated LSP.
In this scenario, leptons and jets originating from on-shell $W, Z, \hsm$-bosons are relatively hard.
Observing excess over the SM backgrounds in the signals with and without soft photons with  $3\ell + \etmiss$ and $1\ell + 2b  + \etmiss$ at the HL-LHC would promptly point back to the scenario that includes singlino-bino co-annihilation with relatively heavier higgsinos. We also point out that, not only in the co-annihilation scenario, a similar signal at the LHC could appear from the singlino-dominated DM, which is mainly annihilated through the $\hsm$ resonance funnel.

It is further pointed out that in the DMDD blind spot regions, higgsino-like neutralinos, being the $\ntrltwo$ and $\ntrlthree$, can exhibit a significant large radiative decay mode into photon in the singlino-higgsino co-annihilation scenario. The bino-like fourth neutralino state indirectly impacts both
collider and DMDD phenomena. It tempers the singlino-like LSP in such a way that the DMDD-SI cross-section remains small and modifies the compositions of various neutralinos in such a way that higgsino-like neutralinos prefer to exhibit radiative photonic decay mode. We observe that in such a compressed scenario, the decaying objects from higgsino-like neutralino ($\ntrltwothree$) and charginos ($\charonepm$) could remain mostly soft from their associated production processes, and it is difficult to suppress the SM backgrounds.
On the contrary, considering an initial state radiation jet associated with the $pp \rightarrow \ntrltwothree \charonepm$ process can be analysed by considering a hard mono-jet with significant missing energy and at least one photon in the signal. We provide the signal cross-sections and kinematic distributions of these objects, which indicate that the process of $pp \rightarrow \ntrltwothree \charonepm$ with an ISR would be a more efficient one to suppress the SM backgrounds.
It would be important for the ATLAS/CMS collaborations to conduct an investigation into this potential signal. Exploring radiative photonic decays of higgsino-like neutralinos through such a search could facilitate a more thorough examination of compressed scenarios and the blind spot regions for DMDD within the NMSSM.
%It holds the potential to discover new particles at the LHC.
%

~\\
\textbf{Acknowledgements} \\
~\\
C.W.\ would like to thank the Aspen Center for Physics, which is supported by National Science Foundation grant No.~PHY-1607611, where part of this work has been done. We would like to thank S.~Baum, M.~Carena, A.~Datta, C.~Hugonie, T.~Ou, D.~Rocha and N.~Shah for useful discussions and comments. S.R. and C.W.\ have been partially supported by the U.S.~Department of Energy under contracts No.\ DEAC02- 06CH11357 at Argonne National Laboratory. The work of C.W.\ at the University of Chicago has also been supported by the DOE grant DE-SC0013642.

\end{document}